\documentclass[twocolumn,trackchanges]{aastex7}
\usepackage{subcaption}

\newcommand{\unit}[1]{\ensuremath{\mathrm{\,#1}}\xspace}

\newcommand{\feh}         {\mbox{[Fe/H]}}
\newcommand{\kms}         {km\,s$^{-1}$}

\newcommand{\msun}        {\unit{M_\odot}}
\newcommand{\lsun}        {\unit{L_\odot}}

\shorttitle{A Transiting Giant Planet Around a Metal-Poor Star}
\shortauthors{J.~D.~Simon et al.}

\begin{document}

\title{TOI-7169~b: A Hot Jupiter Transiting a Metal-Poor Star}

\author[orcid=0000-0002-4733-4994]{Joshua D. Simon}
\affiliation{Observatories of the Carnegie Institution for Science, 813 Santa Barbara St., Pasadena, CA  91101}
\email[show]{jsimon@carnegiescience.edu}  

\author[orcid=0000-0001-8812-0565]{Joseph E. Rodriguez}
\affiliation{Center for Data Intensive and Time Domain Astronomy, Department of
Physics and Astronomy, Michigan State University, East Lansing, MI 48824,
USA}
\email{jrod@msu.edu}

\author[orcid=0000-0001-9261-8366]{Jhon Yana Galarza}
\affiliation{Departamento de Astronomía, Universidad de Concepción, Casilla 160-C, Concepción, Chile}
\email{jyanagalarza@carnegiescience.edu}

\author[orcid=0000-0001-9911-7388]{David W. Latham}
\affiliation{Center for Astrophysics | Harvard \& Smithsonian, 60 Garden St., Cambridge, MA 02138, USA}
\email{dlatham@cfa.harvard.edu}

\author[orcid=0000-0003-0741-7661]{Victoria DiTomasso}
\affiliation{Center for Astrophysics | Harvard \& Smithsonian, 60 Garden St., Cambridge, MA 02138, USA}
\email{victoria.ditomasso@cfa.harvard.edu}

\author[orcid=0000-0001-6588-9574]{Karen A. Collins}
\affiliation{Center for Astrophysics | Harvard \& Smithsonian, 60 Garden St., Cambridge, MA 02138, USA}
\email{karen.collins@cfa.harvard.edu}

\author[orcid=0000-0002-7382-0160]{Jack Schulte}
\affiliation{Center for Data Intensive and Time Domain Astronomy, Department of
Physics and Astronomy, Michigan State University, East Lansing, MI 48824, USA}
\email{jschulte@msu.edu}

\author[orcid=0000-0002-7155-679X]{Anirudh Chiti}
\affiliation{Kavli Institute for Particle Astrophysics \& Cosmology, Stanford University, Stanford, CA 94305, USA}
\affiliation{Brinson Prize Fellow}
\email{achiti@stanford.edu}

\author[orcid=0000-0002-8964-8377]{Samuel N. Quinn}
\affiliation{Center for Astrophysics | Harvard \& Smithsonian, 60 Garden St., Cambridge, MA 02138, USA}
\email{squinn@cfa.harvard.edu}

\author[orcid=0000-0001-9178-3992]{Mohammad K. Mardini}
\affiliation{Department of Physics and Kavli Institute for Astrophysics and Space Research, Massachusetts Institute of Technology, Cambridge, MA 02139, USA}
\email{mardini_mohammad@hotmail.com}

\author[orcid=0000-0001-8401-4300]{Shubham Kanodia}
\affiliation{Earth and Planets Laboratory, Carnegie Institution for Science, 5241 Broad Branch Road NW, Washington, DC 20015, USA}
\email{skanodia@carnegiescience.edu}

\author[orcid=0009-0008-2801-5040]{Johanna K. Teske}
\affiliation{Earth and Planets Laboratory, Carnegie Institution for Science, 5241 Broad Branch Road NW, Washington, DC 20015, USA}
\affiliation{Observatories of the Carnegie Institution for Science, 813 Santa Barbara St., Pasadena, CA  91101}
\email{jteske@carnegiescience.edu}

\author[orcid=0000-0001-6957-1627]{Peter S. Ferguson}
\affiliation{DiRAC Institute, Department of Astronomy, University of Washington, 3910 15th Ave. NE, Seattle, WA, 98195, USA}
\email{pferguso@uw.edu}

\author[orcid=0000-0001-7961-3907]{Samuel W. Yee}
\affiliation{Department of Physics \& Astronomy, University of California Los Angeles, Los Angeles, CA 90095, USA}
\email{syee@astro.ucla.edu}

\author[orcid=0000-0001-5603-6895]{T.~G. Tan}
\affiliation{Perth Exoplanet Survey Telescope, Perth, Western Australia, Australia}
\email{tgtan@bigpond.net.au}

\author[orcid=0000-0001-6253-0179]{Khalid Alsubai}
\affiliation{Hamad bin Khalifa University (HBKU), Qatar Foundation, P.O. Box 5825, Doha, Qatar}
\email{alsubai@yahoo.com}

\author[orcid=0000-0003-1464-9276]{Khalid Barkaoui}
\affiliation{Instituto de Astrof\'{ı}sica de Canarias (IAC), Calle V\'{ı}a L\'{a}ctea s/n, E-38200
La Laguna, Tenerife, Spain}
\affiliation{Astrobiology Research Unit, Universit\'{e} de Li\`{e}ge, 19C All\'{e}e du 6 Ao\^{u}t,
4000 Li\`{e}ge, Belgium}
\affiliation{Department of Earth, Atmospheric and Planetary Science, Massachusetts
Institute of Technology, 77 Massachusetts Avenue, Cambridge, MA 02139,
USA}
\email{khalid.barkaoui@uliege.be}

\author[orcid=0000-0001-6285-9847]{Zouhair Benkhaldoun}
\affiliation{Department of Applied Physics and Astronomy, and Sharjah Academy for
Astronomy, Space Sciences and Technology, University of Sharjah, United Arab Emirates}
\affiliation{Cadi Ayyad University, Oukaimeden Observatory, High Energy Physics,
Astrophysics and Geoscience Laboratory, FSSM, Morocco}
\email{zbenkhaldoun@sharjah.ac.ae}

\author[orcid=0000-0003-4647-7114]{Krzysztof Bernacki}
\affiliation{Silesian University of Technology, Akademicka 16, 44-100 Gliwice, Poland}
\email{Krzysztof.Bernacki@polsl.pl}

\author[]{Jaikrit Bhattacharya}
\affiliation{Hamilton College, 198 College Hill Road, Clinton, NY 13413, USA}
\email{jbhattac@hamilton.edu}

\author[orcid=0000-0002-6424-3410]{Jerome P. de Leon}
\affiliation{Komaba Institute for Science, The University of Tokyo, 3-8-1
Komaba, Meguro, Tokyo 153-8902, Japan}
\email{jpdeleon@g.ecc.u-tokyo.ac.jp}

\author[0009-0002-9833-0667]{Sarah J. Deveny}
\affiliation{Bay Area Environmental Research Institute, Moffett Field, CA 94035, USA}
\affiliation{NASA Ames Research Center, Moffett Field, CA 94035, USA}
\email{deveny@baeri.org}

\author[orcid=0000-0002-0885-7215]{Mark E. Everett}
\affiliation{NSF NOIRLab, 950 N. Cherry Ave., Tucson, AZ 85719, USA}
\email{mark.everett@noirlab.edu}

\author[orcid=0000-0002-9436-2891]{Izuru Fukuda}
\affiliation{Department of Multi-Disciplinary Sciences, Graduate School of Arts and Sciences, The University of Tokyo, 3-8-1 Komaba, Meguro, Tokyo 153-8902, Japan}
\email{izuru-fukuda@g.ecc.u-tokyo.ac.jp}

\author[orcid=0000-0002-4909-5763]{Akihiko Fukui}
\affiliation{Komaba Institute for Science, The University of Tokyo, 3-8-1
Komaba, Meguro, Tokyo 153-8902, Japan}
\affiliation{Instituto de Astrofísica de Canarias (IAC), E-38205 La Laguna, Tenerife, Spain}
\email{afukui@g.ecc.u-tokyo.ac.jp}

\author[orcid=0000-0003-1462-7739]{Micha{\"e}l Gillon}
\affiliation{Astrobiology Research Unit, Universit\'{e} de Li\`{e}ge, 19C All\'{e}e du 6 Ao\^{u}t,
4000 Li\`{e}ge, Belgium}
\email{michael.gillon@uliege.be}

\author[orcid=0000-0002-5463-9980]{Arvind F. Gupta}
\affiliation{NSF NOIRLab, 950 N. Cherry Ave., Tucson, AZ 85719, USA}
\email{arvind.gupta@noirlab.edu}

\author[0000-0002-2532-2853]{Steve~B.~Howell}
\affiliation{NASA Ames Research Center, Moffett Field, CA 94035, USA}
\email{steve.b.howell@nasa.gov}

\author[orcid=0000-0001-8923-488X]{Emmanuel Jehin}
\affiliation{Space Sciences, Technologies and Astrophysics Research (STAR) Institute,
Universit\'{e} de Li\`{e}ge, All\'{e}e du 6 Ao\^{u}t 19C, B-4000 Li\`{e}ge, Belgium}
\email{ejehin@uliege.be}

\author[orcid=0000-0002-4197-7374]{Gaia Lacedelli}
\affiliation{Instituto de Astrofísica de Canarias (IAC), E-38205 La Laguna, Tenerife, Spain}
\affiliation{Departamento de Astrofísica, Universidad de La Laguna (ULL), E-38206 La Laguna, Tenerife, Spain}
\email{gaia.lacedelli@iac.es}

\author[orcid=0009-0009-2881-7112]{Adam Lark}
\affiliation{Hamilton College, 198 College Hill Road, Clinton, NY 13413, USA}
\email{alark@hamilton.edu}

\author[0000-0001-7746-5795]{Colin Littlefield}
\affiliation{Bay Area Environmental Research Institute, Moffett Field, CA 94035, USA}
\affiliation{NASA Ames Research Center, Moffett Field, CA 94035, USA}
\email{littlefield@baeri.org}

\author[orcid=0000-0001-9087-1245]{Felipe Murgas}
\affiliation{Instituto de Astrofísica de Canarias (IAC), E-38205 La Laguna, Tenerife, Spain}
\affiliation{Departamento de Astrofísica, Universidad de La Laguna (ULL), E-38206 La Laguna, Tenerife, Spain}
\email{fmurgas@iac.es}

\author[orcid=0000-0001-8511-2981]{Norio Narita}
\affiliation{Komaba Institute for Science, The University of Tokyo, 3-8-1
Komaba, Meguro, Tokyo 153-8902, Japan}
\affiliation{Instituto de Astrofísica de Canarias (IAC), E-38205 La Laguna, Tenerife, Spain}
\affiliation{Astrobiology Center, 2-21-1 Osawa, Mitaka, Tokyo 181-8588, Japan}
\email{narita@g.ecc.u-tokyo.ac.jp}

\author[orcid=0000-0003-0987-1593]{Enric Palle}
\affiliation{Instituto de Astrofísica de Canarias (IAC), E-38205 La Laguna, Tenerife, Spain}
\affiliation{Departamento de Astrofísica, Universidad de La Laguna (ULL), E-38206 La Laguna, Tenerife, Spain}
\email{epalle@iac.es}

\author[orcid=0000-0001-5519-1391]{Hannu Parviainen}
\affiliation{Instituto de Astrofísica de Canarias (IAC), E-38205 La Laguna, Tenerife, Spain}
\affiliation{Departamento de Astrofísica, Universidad de La Laguna (ULL), E-38206 La Laguna, Tenerife, Spain}
\email{hannu@iac.es}

\author[orcid=0000-0003-3184-5228]{Adam Popowicz}
\affiliation{Silesian University of Technology, Akademicka 16, 44-100 Gliwice, Poland}
\email{adam.popowicz@polsl.pl}

\author[orcid=0000-0001-8227-1020]{Richard P. Schwarz}
\affiliation{Center for Astrophysics | Harvard \& Smithsonian, 60 Garden St., Cambridge, MA 02138, USA}
\email{rpschwarz@comcast.net}

\author[0000-0002-1836-3120]{Avi Shporer}
\affiliation{Department of Physics and Kavli Institute for Astrophysics and Space Research, Massachusetts Institute of Technology, Cambridge, MA 02139, USA}
\email{shporer@mit.edu}

\author[orcid=0000-0002-0345-2147]{Abderahmane Soubkiou}
\affiliation{Astrobiology Research Unit, Universit\'{e} de Li\`{e}ge, 19C All\'{e}e du 6 Ao\^{u}t,
4000 Li\`{e}ge, Belgium}
\email{abdousoubkiou@gmail.com}

\author[orcid=0000-0003-2127-8952]{Francis P. Wilkin}
\affiliation{ Department of Physics and Astronomy, Union College, 807 Union Street, Schenectady, NY 12308, USA}
\email{wilkinf@union.edu}

\begin{abstract}

Most known planets are found around metal-rich host stars, which has made it difficult to determine whether a lower metallicity limit for planet formation exists and how the properties of planets born in low-metallicity environments may differ from those with metal-rich origins.
We present the discovery and characterization of TOI-7169~b (TIC~372048733~b), a hot Jupiter that is orbiting a spectroscopically-confirmed metal-poor ($\feh = -0.72 \pm 0.05$) host star.  Based on photometry from TESS and follow-up ground-based imaging, we measure an orbital period of $3.4373125^{+0.0000020}_{-0.0000019}$~d and a planetary radius of $1.475\pm0.029~R_{{\rm Jup}}$.  We use TRES spectroscopy to determine a mass for TOI-7169~b of $0.41\pm0.14~M_{{\rm Jup}}$.  The planet is therefore inflated, with a low density of $0.159^{+0.055}_{-0.054}$~g~cm$^{-3}$.  We also characterize the host star, showing that TOI-7169 is ancient ($12.3 \pm 0.6$~Gyr) and $\alpha$-enhanced ($[\alpha/\textrm{Fe}] \approx 0.3$), but with a Galactocentric orbit that is confined to the thin disk.  TOI-7169 is perhaps the oldest and most metal-poor star currently known to host a transiting giant planet.  Future transmission spectroscopy probing the atmosphere of TOI-7169~b may provide insight into the effect of metallicity on the physical properties of giant planets.

\end{abstract}


\section{Introduction} 
\label{sec:intro}

It is well known that the planet occurrence rate depends on stellar metallicity, in the sense that metal-rich stars are more likely to host planets \citep[e.g.,][]{fv05}.  Based on large samples of planets and extensive surveys of their host stars, recent studies have begun to refine this picture, showing that the occurrence rates of different types of planets and planetary systems may vary in different ways as a function of metallicity \citep[e.g.,][]{buchhave12,mortier13,brewer18,petigura18,rm23,boley24,vb25}.  However, discovery of exoplanets around old and metal-poor stars has remained quite limited, likely because of both lower occurrence rates and the small number of focused searches \citep[e.g.,][]{sozzetti09,santos11,mortier12,faria16,barbato19,boley24}.  The scarcity of known exoplanets with metal-poor host stars hampers our ability to investigate questions such as how planet properties are related to heavy element abundances and what the minimum metallicity required to form planets is.

The NASA Exoplanet Archive list of planets \citep{christiansen25} includes host stars as metal-poor as $\feh = -2.5$, but the objects at $\feh < -1$ all have metallicities determined photometrically rather than spectroscopically.  Even at $-1 < \feh < -0.7$, the list is dominated by radial velocity planet detections, Kepler planet candidates that have not been confirmed, companions that are likely in the brown dwarf mass range, stars lacking robust metallicity measurements, and combinations thereof.  For planets that were identified by radial velocities and do not transit, the planetary properties that can be measured are very limited. 

Among the sample of metal-poor ($\feh < -0.5$) transiting planets, a number of candidates were discovered by the Kepler mission \citep[e.g.,][]{morton16}.  Most of these are small planets with faint host stars, and nearly all lack reliable metallicities.  A notable object in this class is K2-344~b \citep{deleon21}, which is a hot super-Earth around a late K dwarf host with a photometric metallicity of $\feh = -0.95$.  The lowest-metallicity transiting giant planet hosts are WASP-98 at $\feh = -0.60 \pm 0.19$ \citep{hellier14} and WASP-112 at $\feh = -0.64 \pm 0.15$ \citep{anderson14}.  No other transiting giant planets with host stars at $\feh < -0.5$ have been confirmed.


Taking advantage of the large number of planet candidates identified by the Transiting Exoplanet Survey Satellite (TESS) and its nearly all-sky coverage \citep{tess}, we are conducting a new search aimed at identifying planets orbiting low-metallicity host stars.  This effort may help in establishing the lowest metallicity at which the formation of different types of planets is possible, as well as providing better targets for atmospheric characterization of planets that formed in a low-metallicity environment.  In this paper, we present the discovery of TOI-7169~b, a giant planet in a short-period orbit around a metal-poor host.

In \S~\ref{sec:search}, we briefly introduce our search strategy.  We
present our observations of TOI-7169 in \S~\ref{sec:observations}.  In
\S~\ref{sec:analysis}, we describe our measurements of the properties
of the planet and the host star.  We discuss the properties of
TOI-7169~b in the context of the exoplanet population and the Milky
Way in \S~\ref{sec:discussion} and summarize our conclusions in \S~\ref{sec:conclusions}.

\section{Identification of Candidates}
\label{sec:search}

The starting point for our search for planets with metal-poor host stars is the catalog of TESS Objects of Interest (TOIs), which currently includes more than 7000 candidate transiting planets.  Unlike Kepler, TESS is a magnitude-limited survey that is not biased against metal-poor stars.  Some previous studies have used catalogs of metal-poor stars to conduct their own searches for transits in TESS light curves \citep[e.g.,][]{boley21,boley24}.  Here, we take the simpler approach of identifying metal-poor stars that already have transits detected by the TESS team.

We select candidate planets that may be hosted by metal-poor ($\feh <
-0.5$) stars based on several analyses of all-sky, low-resolution
spectrophotometric data (i.e., XP spectra;
\citealt{gaiadr3montegriffo}) from the third data release (DR3) of the
Gaia mission \citep{gaia16a,gaiadr3vallenari}.  The Gaia team provided
a python toolkit to derive photometry in various passbands from this
data product\footnote{https://gaia-dpci.github.io/GaiaXPy-website/}
\citep{gaiaxpyv100}.  We derive photometry in bands that are known to
map onto stellar metallicity, including the SkyMapper filter set
(i.e., $v$, $g$, $i$; \citealt{keller07}) and a narrow-band filter
covering the prominent \ion{Ca}{2} H and K lines
\citep[e.g.,][]{starkenburg17}. Then, we adopt the techniques and
model grid described by \citet{chiti20,chiti21} to derive
metallicities for stars from this Gaia XP-based photometry via
comparison to a grid of synthetic, forward-modeled photometry of stars
spanning a range of effective temperatures, surface gravities, and
metallicities.  The catalog and further details of this analysis will
be presented in Mardini et al. (in preparation).  Separately, the
Pristine collaboration also released a metallicity catalog of sources
in the Gaia XP catalog \citep{martin24}, which we use for our
selection as well.  We cross-match both metallicity catalogs with the set of allTOIs to construct an initial candidate sample.  This selection is independent of kinematics, so it should include low metallicity stars regardless of their association with the thin disk, thick disk, or halo of the Milky Way.

For these candidates, we search for spectroscopic metallicities in the GALAH, APOGEE, and LAMOST surveys to confirm their metal-poor nature.  Most candidates do not have independent metallicities available, but some can be eliminated in this way.  Because the binary fraction of metal-poor stars is high \citep{moe19}, we next check ExoFOP\footnote{https://exofop.ipac.caltech.edu/tess/} and the Gaia DR3 catalog \citep{gaiadr3vallenari} for evidence of binarity.  We discard targets for which speckle or adaptive optics imaging has revealed a close companion.  For stars lacking such high-resolution follow-up imaging, close neighbors in the Gaia stellar catalog, a high reduced unit weight error \citep[RUWE;][]{belokurov20,penoyre22,gaiadr3arenou}, or an ipd\_frac\_multi\_peak value above 2 \citep{gaiaedr3fabricius,gaiaedr3smart,elbadry21} may indicate binarity.  Although a binary companion does not necessarily mean that the TESS signal is of non-planetary origin, it significantly complicates the follow-up, making it challenging to determine which star hosts the transit and potentially biasing precision radial velocity measurements.  We therefore set aside binaries for the time being, which eliminates $\sim2/3$ of the $\sim100$ metal-poor candidates.  After removing confirmed or likely binaries, stars with spectroscopic metallicities demonstrating that they are not metal-poor, and companions larger than $\sim2$~$R_{\textrm{Jup}}$, we are left with 13 candidates with estimated metallicities $-2 < \feh < -0.5$ warranting follow-up observations.

One object that passes this selection process is TOI-7169
(Gaia~DR3~1913763798475865344), which has an estimated metallicity of
$\feh = -0.64$.  The Gaia astrometry contains no indication of an
unresolved companion ($\mathrm{RUWE}=0.94$,
$\mathrm{ipd\_frac\_multi\_peak} = 0$), and the closest neighbor in
the Gaia DR3 catalog is 2\farcs4 away and 6~mag fainter (with a very
different parallax and proper motion, indicating that the two stars
are not physically associated).  However, the Gaia spectroscopy does
reveal a relatively large velocity amplitude of 23~\kms.

TOI-7169 was announced by the TESS collaboration as a TESS Object of Interest (TOI) on 2024 November 14 \citep{guerrero21}.  A transiting signal was detected with the SPOC pipeline in data from sector 83 with a period of 3.44~d, a planetary radius of 15.0~R$_{\oplus}$, and a signal-to-noise ratio of 26.  The basic photometric and astrometric parameters of TOI-7169 are listed in Table~\ref{tab:photom}.

\begin{deluxetable*}{ll}
\tablecolumns{2}
\tabletypesize{\footnotesize}
\tablecaption{\label{tab:photom}TOI-7169 Astrometry and Photometry}
\tablehead{ Parameter & Value }
\startdata
R.A. (hours)  & 23:24:$13.126941 \pm 0.00009$~s \\
Decl. (degrees) & 36:49:$03.24207 \pm 0\farcs00008$ \\
Parallax (mas)  & $2.207 \pm 0.012$ \\
$\mu_{\alpha}\cos{\delta}$ (mas~yr$^{-1}$) & $34.280 \pm 0.011$ \\
$\mu_{\delta}$ (mas~yr$^{-1}$) & $8.358 \pm 0.011$ \\
$BP$ (mag) & $12.717 \pm 0.003$ \\
$G$ (mag) & $12.362 \pm 0.003$ \\
$RP$ (mag) & $11.835 \pm 0.004$ \\
$J$ (mag) & $11.197 \pm 0.021$ \\
$H$ (mag) & $10.873 \pm 0.028$ \\
$K_{\mathrm{S}}$ (mag) & $10.808 \pm 0.018$ \\
$W1$ (mag) & $10.728 \pm 0.023$ \\
$W2$ (mag) & $10.765 \pm 0.020$ \\
$W3$ (mag) & $10.699 \pm 0.087$ \\
$W4$ (mag) & $9.254 \pm 0.526$ \\
\enddata
\tablecomments{Astrometry is taken from Gaia DR3
  \citep{gaiaedr3lindegren}.  Photometry is from Gaia DR3
  \citep{riello21}, 2MASS \citep{cutri03}, and AllWISE \citep{cutri14}.}
\end{deluxetable*}

\section{Observations}
\label{sec:observations}

\subsection{TESS}

TOI-7169 was observed by TESS during its prime mission and the first two extended missions in Sectors 16, 57, 83, and 84.  The TESS observations were taken as 1800~s cadence full-frame images in Sector 16 and 200~s full-frame images in Sectors 57 and 84. In Sector 83, TOI-7169 was one of 12,000 stars selected for observations at 120~s cadence.  Light curves were produced by the Quick Look Pipeline \citep{huang20a,huang20b} for Sectors 16, 57, and 84 and the SPOC pipeline \citep{jenkins16,caldwell20} for Sector 83.  Because the star is fainter than $T=10.5$, it was not classified as a TOI by the Quick Look Pipeline during the TESS prime mission \citep{guerrero21}.  However, the transits are deep enough to be clearly detectable in all sectors, as illustrated below.

\begin{figure}
    \centering
    \includegraphics[width=0.48\textwidth]
    {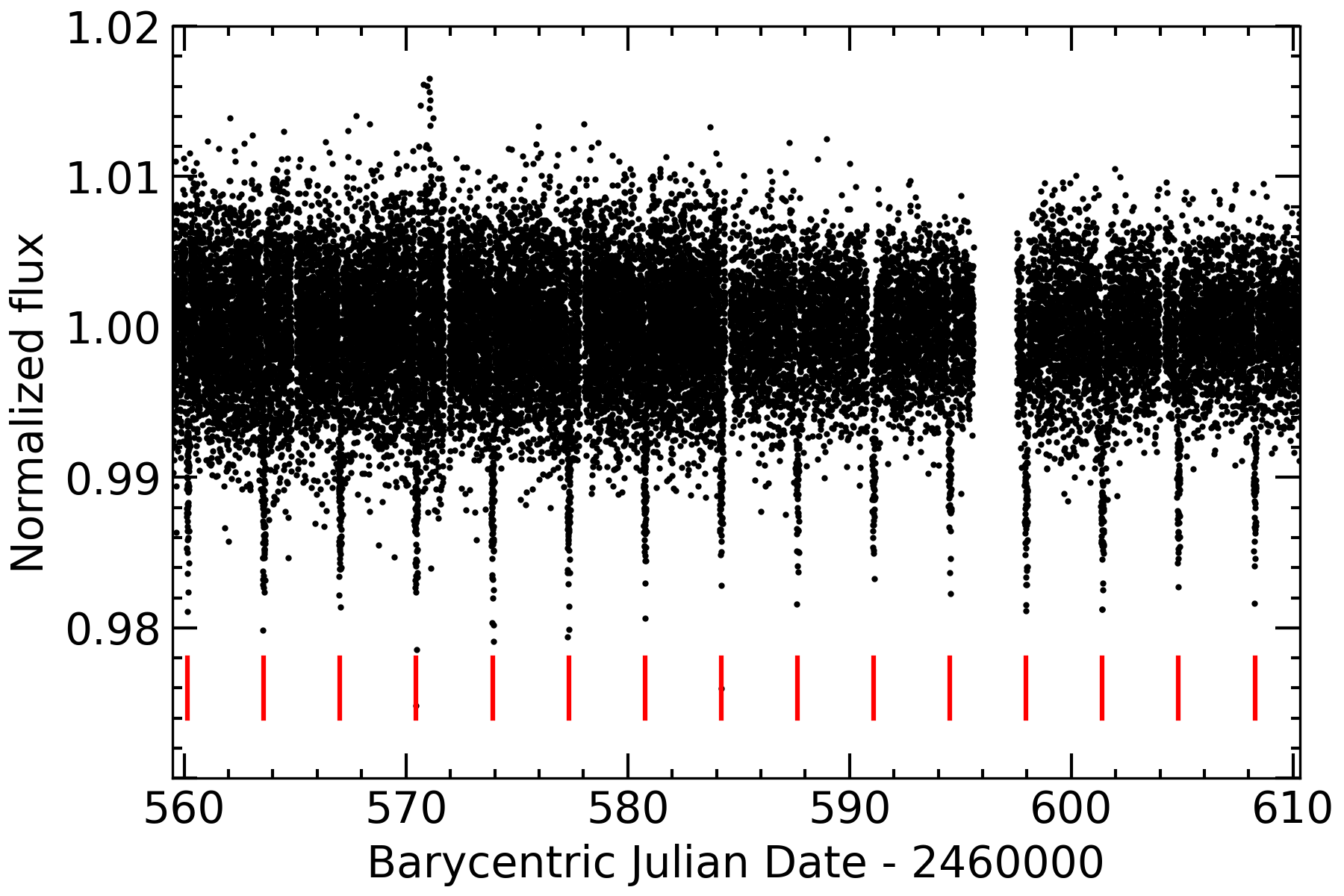}
    \caption{Flattened TESS light curve of TOI-7169 from Sectors 83 ($\mathrm{BJD} < 2460584.5$) and 84 ($\mathrm{BJD} > 2460584.5$).  Sector 83 used an integration time of 120~s and the Sector 84 integration time was 200~s.  The 15 transits of TOI-7169~b that occur across these two sectors are indicated by red hash marks.}\label{fig:tesslc}
\end{figure}

We downloaded the TOI-7169 light curves from the Mikulski Archive for
Space Telescopes (MAST\footnote{https://mast.stsci.edu/}) using the
\texttt{lightkurve} package \citep{lightkurve}.  We processed the
light curves to remove flux variations from stellar activity and
instrumental effects using the spline-fitting tool
\texttt{keplersplinev2} \citep{Vanderburg:2014}, following the
procedures described by \citet{schulte24,schulte25}.  The flattened light curve from Sectors 83 and 84 is illustrated in Fig.~\ref{fig:tesslc}.

\subsection{Ground-based Photometry}

The TESS pixel scale is $\sim 21\arcsec$ pixel$^{-1}$ and photometric apertures typically extend out to roughly 1 arcminute, generally causing multiple stars to blend in the TESS photometric aperture. To determine the true source of the TESS detection and subsequently, transit depth achromaticity, we acquired ground-based time-series follow-up photometry of the field around TOI-7169 as part of the TESS Follow-up Observing Program Sub Group 1 \citep[TFOP SG1;][]{collins:2019}\footnote{https://tess.mit.edu/followup}. We used the {\tt TESS Transit Finder}, which is a customized version of the {\tt Tapir} software package \citep{Jensen:2013}, to schedule our transit observations. We used {\tt AstroImageJ} \citep{Collins:2017} to extract all differential photometric data, except in the case of MuSCAT2 (see below).

As mentioned in \S~\ref{sec:search}, TIC~2040537885 is a close neighbor of TOI-7169, at a separation of $2\farcs4$ and a magnitude difference of $\Delta G = 6.03$ ($\Delta T = 6.01$).  The colors of the two stars are similar, so in an aperture containing both stars, TIC~2040537885 will contribute $\approx3.9$~ppt of the flux in any optical band.  Since the transit depth is 11~ppt (Table~\ref{tab:transitpars}), TIC~2040537885 cannot be responsible for the transit signal.  However, the photometric apertures used for both TESS and all observations discussed below include both stars.

Four partial transits of TOI-7169 were initially observed. The first transit observation was obtained with KeplerCam on UTC 2025 June 22 in Sloan $i'$ band.  A second transit was observed with the TRAPPIST-North-0.6m \citep{Barkaoui2019_TN} telescope in B band on 2025 July 12.  The following transit on 2025 July 16 was observed in Sloan $g'$ and Sloan $i'$ bands from the Las Cumbres Observatory Global Telescope \citep[LCOGT;][]{Brown13} 1.0\,m network node at McDonald Observatory near Fort Davis, Texas, United States. The 1.0\,m telescopes are equipped with $4096\times4096$ SINISTRO cameras providing a plate scale of $0\farcs389$~pixel$^{-1}$ and a field of view of $26\arcmin\times26\arcmin$. All LCOGT images were calibrated by the standard LCOGT {\tt BANZAI} pipeline \citep{McCully:2018}. Another transit was observed with the same filters and telescope on 2025 August 23. These data confirmed that the transit is associated with TOI-7169 rather than a nearby star and updated the period slightly to $P = 3.4373111$~d.

The first full transit observation was obtained on 2025 August 29 with the MuSCAT2 multi-color imager \citep{narita19} on the 1.52~m Telescopio Carlos S\'{a}nchez (TCS).  MuSCAT2 is capable of simultaneous imaging in four bands, each using a 1k$\times$1k CCD camera with a pixel scale of $0\farcs44$ pixel$^{-1}$ and a field of view of $7\farcm4 \times 7\farcm4$.
We used the MuSCAT2 pipeline\footnote{https://github.com/hpparvi/MuSCAT2\_transit\_pipeline}, described in \citet{Parviainen2020}, to reduce the data and carry out aperture photometry over a set of comparison stars and aperture sizes.  The photometry was contaminated only by TIC~2040537885, but again this object is too faint to be the sourece of the transit signal.
The optimal light curves were determined through global optimization of a model consisting of the five brightest comparison stars and contaminated aperture radii smaller than 9\farcs6, while the transit and baseline variations were simultaneously modeled using a linear combination of covariates. As reported in TFOP SG1, we detected an ~on-time transit with depths in the four bands of $(R_{p}/R_{*})^{2} =  9.8, 10.5, 11.1, 10.0$ ppt, which confirms the depth is achromatic within the uncertainties.

A second full transit window of TOI-7169 was observed with a Sloan $r'$ filter on UTC 16 September 2025 from Hamilton College Observatory in Clinton, NY, USA. The 0.51~m telescope is equipped with a QHY600M detector, which has an image scale of 0\farcs45~pixel$^{-1}$, resulting in a $35\farcm9 \times 24\farcm0$ field of view.  We extracted the differential photometric data with circular photometric apertures having radius $4\farcs05$, which excluded the flux from the nearest known neighbor other than TIC~2040537885.

A third full transit of TOI-7169 was observed with the LCOGT 1.0\,m network node at Teide Observatory on the island of Tenerife on UT 2025 September 29. The observations were carried out simultaneously in the Sloan $i'$ and Sloan $g'$ filters with exposure times of 28 and 33\,s, respectively. We extracted photometry using circular $3\farcs5$ photometric apertures.
  
The full transit observations are listed in Table~\ref{tab:transitpars}, and lightcurves from these transits are included in the EXOFASTv2 analysis described in \S~\ref{sec:exofast}.  

\begin{deluxetable*}{lrcc}
\tablecaption{\label{tab:transitpars}Median values and 68\% confidence intervals for transit times, impact parameters, and depths}
\tablehead{\colhead{Transit} & \colhead{Epoch} & \colhead{$T_T$} & \colhead{Depth}}
\startdata
TESS UT 2019-09-12 (TESS)   & -526 & $2458752.12400^{+0.00110}_{-0.00120}$    & $0.011104^{+0.000087}_{-0.000089}$ \\
TESS UT 2022-09-30 (TESS)   & -201 & $2459869.25060^{+0.00054}_{-0.00057}$ & $0.011104^{+0.000087}_{-0.000089}$ \\
TESS UT 2024-09-05 (TESS)   & 3    & $2460570.46235 \pm 0.00021$         & $0.011104^{+0.000087}_{-0.000089}$ \\
TESS UT 2024-10-01 (TESS)   & 11   & $2460597.96085 \pm 0.00020$         & $0.011104^{+0.000087}_{-0.000089}$ \\
MuSCAT2 UT 2025-08-30 (g')  & 104  & $2460917.63092 \pm 0.00017$         & $0.011544^{+0.000088}_{-0.000086}$ \\
MuSCAT2 UT 2025-08-30 (i')  & 104  & $2460917.63092 \pm 0.00017$         & $0.011165 \pm 0.000093$          \\
MuSCAT2 UT 2025-08-30 (r')  & 104  & $2460917.63092 \pm 0.00017$         & $0.011316^{+0.000082}_{-0.000084}$ \\
MuSCAT2 UT 2025-08-30 (z')  & 104  & $2460917.63092 \pm 0.00017$         & $0.010965^{+0.000090}_{-0.000091}$ \\
Hamilton UT 2025-09-15 (r') & 109  & $2460934.81748 \pm 0.00018$         & $0.011316^{+0.000082}_{-0.000084}$ \\
LCO-Teid UT 2025-09-29 (g') & 113  & $2460948.56673 \pm 0.00018$         & $0.011544^{+0.000088}_{-0.000086}$ \\
LCO-Teid UT 2025-09-29 (i') & 113  & $2460948.56673 \pm 0.00018$         & $0.011165 \pm 0.000093$          \\
\enddata
\end{deluxetable*}

\subsection{Speckle Imaging}

\begin{figure*}
    \centering
    \begin{subfigure}[b]{0.47\textwidth}            
            \includegraphics[width=\textwidth]{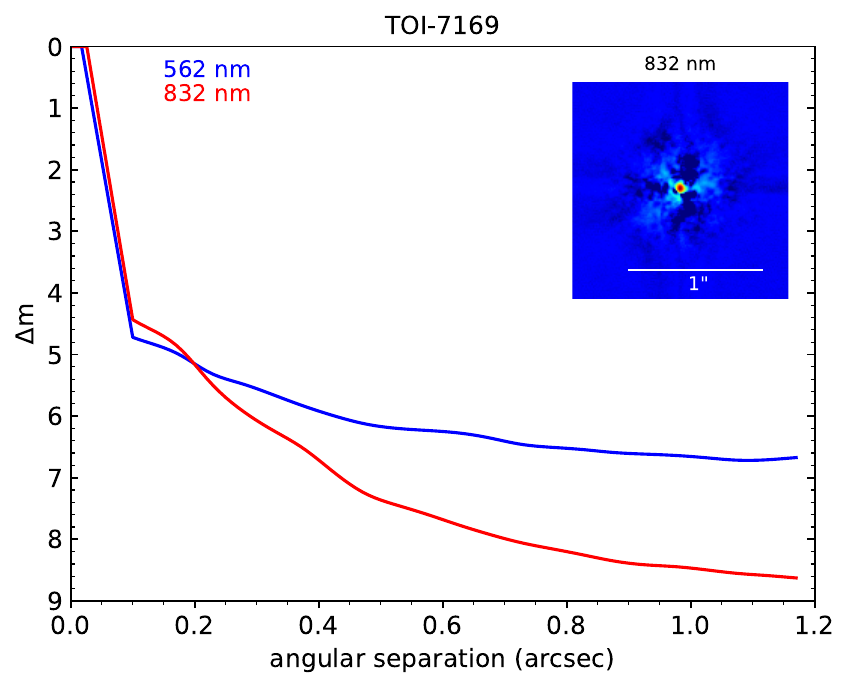}
    \end{subfigure}%
    \begin{subfigure}[b]{0.47\textwidth}
            \centering
            \includegraphics[width=\textwidth]{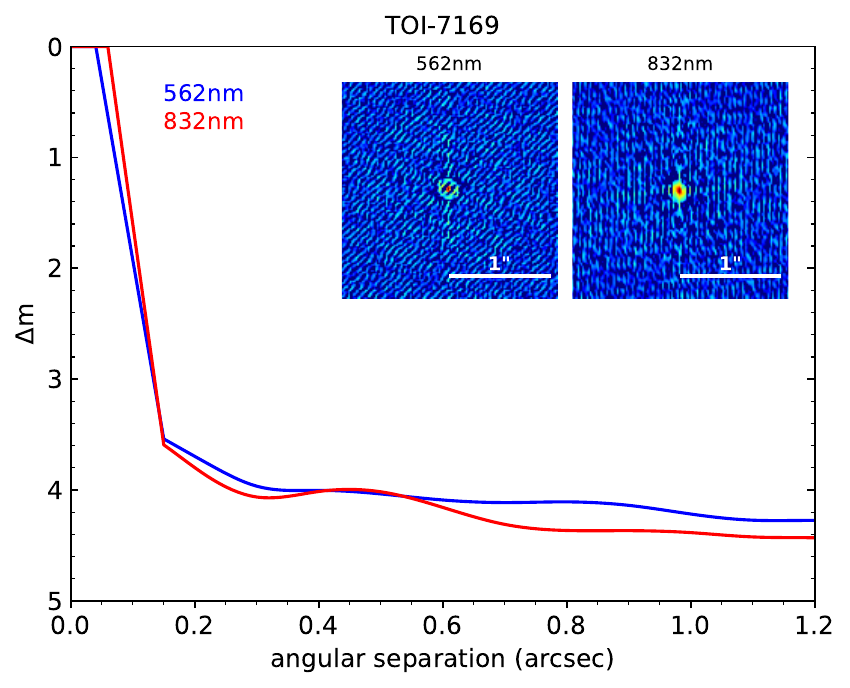}
    \end{subfigure}
    \caption{Results from speckle imaging of TOI-7169. (Left) Gemini/'Alopeke detection limits.  The blue and red curves show the $5\sigma$ contrast limits in magnitudes reached in the 562~nm and 832~nm filters, respectively.  The inset in the upper right presents the 832~nm image of the field, with no sources other than the host star detected. (Right) WIYN/NESSI $5\sigma$ detection limits.  The blue and red curves show the contrast limits reached in the 562~nm and 832~nm filters and the insets show the two reconstructed images.}\label{fig:speckle}
\end{figure*}

We obtained high-resolution speckle imaging of TOI-7169 with the 'Alopeke instrument \citep{scott21} on the Gemini North telescope on 2025 August 6.  We observed the star with two narrow-band filters, one centered at 562~nm and the other at 832~nm.  The data were reduced following the procedures described by \citet{howell11}.  No nearby sources were detected in either filter.  We set lower limits on the magnitude difference of potential companions to TOI-7169 of $\sim4.5$~mag at a separation of $\sim0\farcs1$ (Fig.~\ref{fig:speckle}).  At larger separations, the contrast limit reaches $\sim6.6$~mag in the bluer filter and $\sim8.6$~mag in the redder filter.  Given the observed transit depth, the possibility of the transit actually occurring on a companion star is ruled out for separations between 0\farcs1 and 1\farcs2.

We also observed TOI-7169 with the NN-Explore Exoplanet Stellar Speckle Imager \citep[NESSI;][]{scott18} on the WIYN 3.5\,m telescope\footnote{The WIYN Observatory is a joint facility of the University of Wisconsin-Madison, Indiana University, NSF NOIRLab, the Pennsylvania State
University, and Princeton University.} at Kitt Peak National Observatory. Simultaneous sets of 1000 40 ms diffraction-limited frames were collected on the night of 2025 December 9 in the 562 nm and 832 nm filters using the blue and red cameras, respectively. The speckle images were reconstructed following the methods outlined in \citet{howell11}. We find no evidence for any nearby sources; the achieved $5\sigma$ contrast limits are sufficient to rule out resolved sources as faint as $\Delta m_{562}=4.0$ and $\Delta m_{832}=4.0$ at 0\farcs5 and $\Delta m_{562}=4.2$ and $\Delta m_{832}=4.4$ at 1\farcs0.

\subsection{Spectroscopy}

As part of both our initial vetting of the transit signal and later follow-up to determine the planet mass, we obtained 19 spectra of TOI-7169 with the Tillinghast Reflector Echelle Spectrograph (TRES) on the 1.5~m telescope at Whipple Observatory \citep{sf07,furesz08}.  Most of the spectra are concentrated between 2024 December 15 and 2025 February 5, but two observations were obtained in 2016 September and two were obtained in 2025 October.  TRES provides high-resolution ($R = 44000$) spectra extending from 3850--9100~\AA.  Data reduction followed the procedures described by \citet{buchhave2010} and \citet{quinn12}.  The signal-to-noise ratio (S/N) of individual spectra was $\sim30$ per resolution element.

We used the TRES spectra to obtain radial velocity measurements of the star's reflex motion resulting from the planetary orbit.  The typical velocity precision obtained from each spectrum was $\sim50$~m~s$^{-1}$, which is sufficient for characterizing a giant planet in a short-period orbit.  We list the measured RVs in Table~\ref{tab:vels}.  

\begin{deluxetable*}{lcc}
\tablecolumns{3}
\tabletypesize{\footnotesize}
\tablecaption{\label{tab:vels}TOI-7169 Radial Velocities}
\tablehead{ Barycentric Julian Date & Relative Velocity & Velocity Uncertainty \\ 
 & [m s$^{-1}$] & [m s$^{-1}$] }
\startdata
2457640.891134 & \phs\phn26.3 & 43.2 \\
2457642.868659 & \phn$-$97.7  & 39.4 \\
2460660.587476 & $-$228.9     & 41.3 \\
2460665.729905 & \phs\phn36.8 & 40.1 \\
2460666.719977 & \phs\phn10.0 & 40.4 \\
2460668.675150 & \phn$-$80.1  & 25.2 \\
2460671.631793 & \phn$-$91.8  & 39.8 \\
2460675.689780 & \phn$-$61.9  & 32.9 \\
2460676.728400 & \phn$-$92.5  & 51.8 \\
2460679.594789 & \phs\phn54.2 & 52.7 \\
2460680.573484 & \phn$-$40.5  & 40.6 \\
2460681.575481 & $-$156.8     & 48.5 \\
2460707.592292 & \phn$-$46.5  & 48.0 \\
2460708.597809 & \phn$-$55.9  & 61.1 \\
2460710.586763 & \phs\phn10.8 & 34.6 \\
2460711.591263 & \phn$-$11.9  & 39.0 \\
2460712.606123 & $-$126.9     & 54.1 \\
2460951.837087 & \phn$-$82.0  & 41.0 \\
2460976.889632 & \phs\phn\phn7.3 & 58.0
%
\enddata
\end{deluxetable*}

\begin{figure*}[t!]
    \centering
    \includegraphics[width=1.0\textwidth]
    {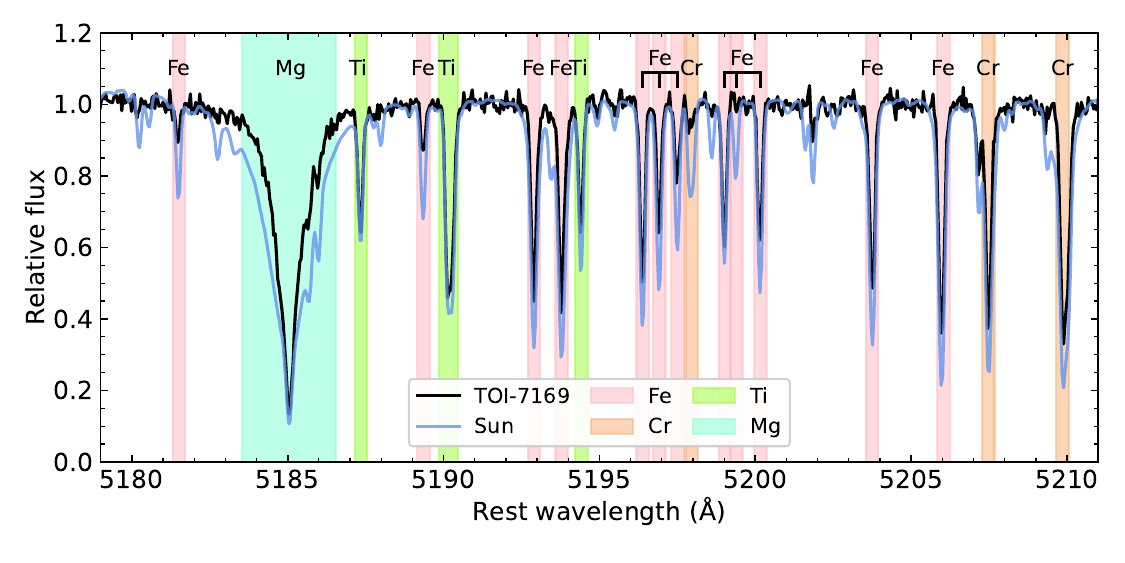}
    \caption{Comparison between the spectrum of TOI-7169 (black) and the Sun (blue).  The temperatures of the two stars are quite similar, so the differences in line strength are primarily due to abundance differences (with the exception of the very strong Mg line at 5185~\AA).  Spectral lines of iron-peak (Fe and Cr) and $\alpha$ elements (Mg and Ti) are shaded as indicated in the legend. }
    \label{fig:spec}
\end{figure*}

Because of the relatively low S/N of the individual epochs, we co-added the 15 spectra from the end of 2024/beginning of 2025 together to create a single higher S/N spectrum for determining the properties of the host star. We used \textsc{IRAF\footnote{\textsc{IRAF} is distributed by the National Optical Astronomical Observatories, which is operated by the Association of Universities for Research in Astronomy, Inc., under a cooperative agreement with the National Science Foundation.}} to apply barycentric and RV corrections with the \textsc{dopcor} task. Each spectrum was then normalized in pixel space using the \textsc{continuum} task with second- to fifth-order polynomials. Finally, the normalized spectra were combined using the \textsc{scombine} task. The resulting spectrum has a S/N of $\sim80$ at $\sim630$ nm. A small section of the TRES spectrum is displayed in Fig.~\ref{fig:spec}, with a TRES solar spectrum (created from the coaddition of all available TRES asteroid spectra by \citealt{pass25}) overplotted for comparison.

\section{Analysis}
\label{sec:analysis}

\subsection{Host Star Characterization}
\label{sec:hoststar}


We took two independent approaches to measuring the stellar parameters of TOI-7169.  First, we used the uberMS framework introduced by \citet{pass25} to fit the order of the TRES data containing the Mg triplet.  UberMS employs a neural network to compare the stellar spectrum and photometry with synthetic spectra and theoretical isochrones.  The photometric analysis includes fluxes from Gaia \citep{riello21}, 2MASS \citep{cutri03}, and WISE \citep{wright10}, along with the MIST isochrones \citep{choi16}.  We made two minor modifications to the procedures adopted by \citet{pass25} and \citet{ditomasso25}: (1) we widened the prior on the extinction to encompass from 0 to $1.5\times$ the value from the \citet{sfd98} dust map, and (2) we added WISE W1 and W2 photometry to better constrain the extinction.

The uberMS fit finds the following parameters for TOI-7169: $T_{\mathrm{eff}} = 5840 \pm 100$~K, $\log{g} = 4.05 \pm 0.09$, $v_{micro} = 1.18$~\kms, $\feh = -0.68 \pm 0.04$, and $[\alpha/{\textrm{Fe}}] = 0.22 \pm 0.03$ \citep{ditomasso26}.  The derived distance to the star is 454~pc and the extinction is $A_{V} = 0.355$~mag, consistent with the Gaia DR3 parallax and the 3D dust map from \citet{green19}.

Second, we carried out a traditional equivalent width (EW) analysis \citep[e.g.,][]{yg21,yg24} of the coadded TRES spectrum. We measured EWs by fitting Gaussian profiles with the \textsc{KAPTEYN} \texttt{kmpfit} Python package \citep{KapteynPackage}, using pseudo-continuum windows of about 6 \AA. All measurements were performed manually, line by line, including careful deblending of blended features. We adopted the line list from \citet{yg24}, which is based on the line list of \citet{Melendez:2014ApJ...791...14M} and provided atomic data such as statistical weights and oscillator strengths. We computed the ionization and excitation balance from 62 \ion{Fe}{1} lines and 15 \ion{Fe}{2} lines using the Qoyllur-quipu ($\mathrm{q}^{2}$) Python code \citep{Ramirez:2014A&A...572A..48R}\footnote{\url{https://github.com/astroChasqui/q2_tutorial}}, which is configured to employ the Kurucz \textsc{ODFNEW} model atmospheres \citep{Castelli:2003IAUS..210P.A20C} and the 2019 version of the local thermodynamic equilibrium (LTE) radiative transfer code \textsc{MOOG} \citep{Sneden:1973PhDT.......180S}. We adopted the \citet{asplund21} solar abundances.  The EW-based stellar parameters favor a slightly lower temperature than the uberMS fit, but the surface gravity and metallicity are consistent: $T_{\mathrm{eff}} = 5714 \pm 42$~K, $\log{g} = 3.913 \pm 0.130$, $v_{\mathrm{micro}} = 1.18 \pm 0.08$~\kms, and $\feh = -0.72 \pm 0.05$.
The star is a slightly evolved early G star, with a radius of $1.50\pm0.03~R_{\odot}$. 
Taking the $\alpha$-element abundance as an average of [Mg/Fe], [Si/Fe], and [Ca/Fe], we find $[\alpha/{\textrm{Fe}}] = 0.29 \pm 0.04$.  We adopt the stellar parameters from the EW-based analysis for the remainder of the paper (see Table~\ref{tab:stellarparams}).

\begin{figure}
    \centering
    \includegraphics[width=0.48\textwidth]
    {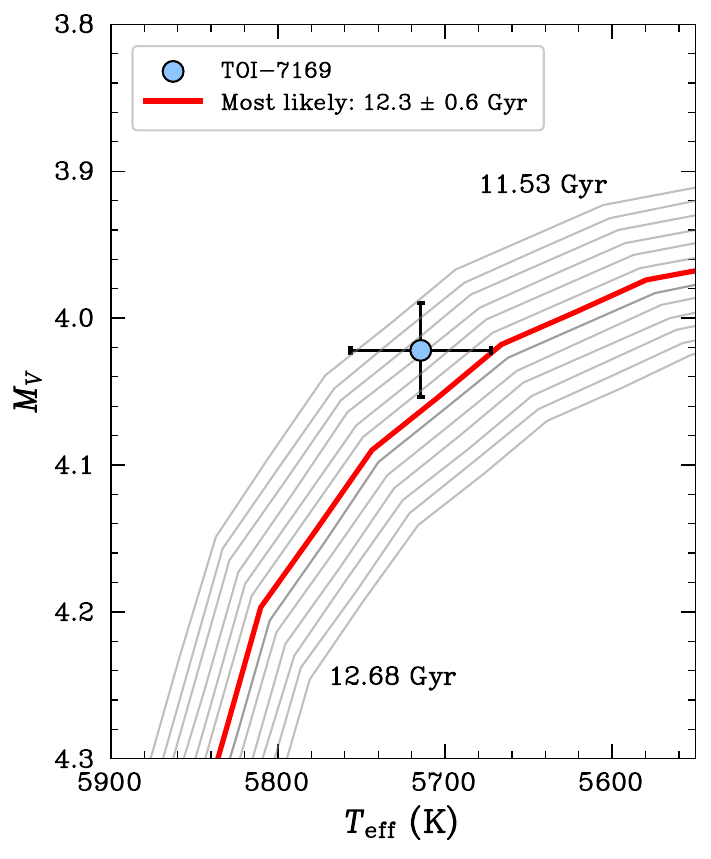}
    \caption{Derived absolute magnitude and effective temperature of TOI-7169 from the q$^{2}$ package, compared to Yonsei-Yale isochrones.  The red isochrone represents the best-fitting age, which does not exactly match the derived stellar properties because it is a marginalized Bayesian posterior in a higher-dimensional parameter space.  The tight constraints on both parameters, and hence on the age, are consequences of the small error bars on the stellar parallax and $G$-band magnitude from Gaia.
    }
    \label{fig:age}
\end{figure}

The stellar age, mass, and radius were also inferred using the
$\mathrm{q}^{2}$ package. Briefly, this Bayesian method adopts the
Yonsei–Yale stellar evolution isochrones
\citep{Yi:2001ApJS..136..417Y, Demarque:2004ApJS..155..667D} and
combines spectroscopically inferred stellar parameters with Gaia
magnitudes, parallaxes, and reddening to construct the probability
distribution functions for the stellar age, mass, and radius. As the
star is enhanced in $\alpha$-elements, we applied an offset to correct
their contribution to the global metallicity, following Equation (1)
of \citet{Salaris:1993ApJ...414..580S}. Further details are provided
in \citet{Ramirez:2014A&A...572A..48R},
\citet{Yana_Galarza:2021MNRAS.504.1873Y}, and
\citet{2023MNRAS.522.3217M}.  We find that the age of TOI-7169 is
$12.3 \pm 0.6$~Gyr (see Fig. \ref{fig:age}) and the mass is $0.885 \pm
0.010~M_{\odot}$. This approach results in small statistical uncertainties because of the high precision of the parallax ($2.207 \pm 0.012$~mas~yr$^{-1}$) and apparent magnitude ($G = 12.362 \pm 0.003$).  In general, absolute ages for old stars determined from isochrones are model-dependent at the $\sim1$~Gyr level \citep[e.g.,][]{brown14}, although in this case we find identical results with the Dartmouth isochrones \citep{dotter08} as well.  Chemical clocks such as the [Y/Mg] abundance ratio\footnote{The calibration from \citet{shejeelammal24} is only defined down to $\feh = -0.71$, but given the negligible difference between that value and the derived metallicity of TOI-7169, we assume that no extrapolation is needed.} \citep{shejeelammal24} provide a consistent age of $14.6 \pm 3.4$~Gyr, but with significantly lower precision.  The derived age is also consistent with the age of 12~Gyr from the thick disk age-metallicity relation of \citet{xr22}, as shown by \citet{adamsredai25} for the similar star TOI-7019.

\begin{deluxetable*}{lccc}
\tablecolumns{4}
\tabletypesize{\footnotesize}
\tablecaption{\label{tab:stellarparams}Adopted Stellar Parameters of TOI-7169}
\tablehead{ $T_{\mathrm{eff}}$ (K) & $\log{g}$ & [Fe/H] & Age (Gyr) }
\startdata
$5714 \pm 42$ & $3.913 \pm 0.130$ & $-0.72 \pm 0.05$ & $12.3 \pm 0.6$
\enddata
\end{deluxetable*}

We derived elemental abundances for \ion{C}{1}, \ion{O}{1}, \ion{Na}{1}, \ion{Mg}{1}, \ion{Al}{1}, \ion{Si}{1}, \ion{Ca}{1}, \ion{Sc}{1}, \ion{Sc}{2}, \ion{Ti}{1}, \ion{Ti}{2}, \ion{V}{1}, \ion{Cr}{1}, \ion{Cr}{2}, \ion{Mn}{1}, \ion{Fe}{1}, \ion{Fe}{2}, \ion{Co}{1}, \ion{Ni}{1}, \ion{Cu}{1}, \ion{Zn}{1}, \ion{Sr}{1}, \ion{Y}{2}, \ion{Ba}{2}, \ion{Ce}{2}, \ion{Nd}{2}, and \ion{Sm}{2}. We obtained oxygen abundances from the high-excitation \ion{O}{1} $\lambda777$ nm triplet and corrected for non-local thermodynamic equilibrium effects using the grids of \citet{amarsi15}. All abundances were estimated from equivalent width measurements, following the same procedure used for \ion{Fe}{1} and \ion{Fe}{2}. We accounted for hyperfine structure and isotopic splitting for \ion{Sc}{1}, \ion{Sc}{2}, \ion{V}{1}, \ion{Mn}{1}, \ion{Co}{1}, \ion{Cu}{1}, \ion{Y}{2}, \ion{Ba}{2}, and \ion{Eu}{2}, adopting data from \cite{McWilliam:1998AJ....115.1640M}, \citet{Prochaska2000a}, \citet{Prochaska2000b}, \citet{klose2002}, \citet{Cohen:2003ApJ...588.1082C}, \citet{Blackwell-Whitehead2005a,Blackwell-Whitehead2005b}, \citet{Lawler2014} and from the Kurucz\footnote{\url{http://kurucz.harvard.edu/linelists.html}} line lists. 

\begin{figure*}
    \centering
    \includegraphics[width=0.995\textwidth]{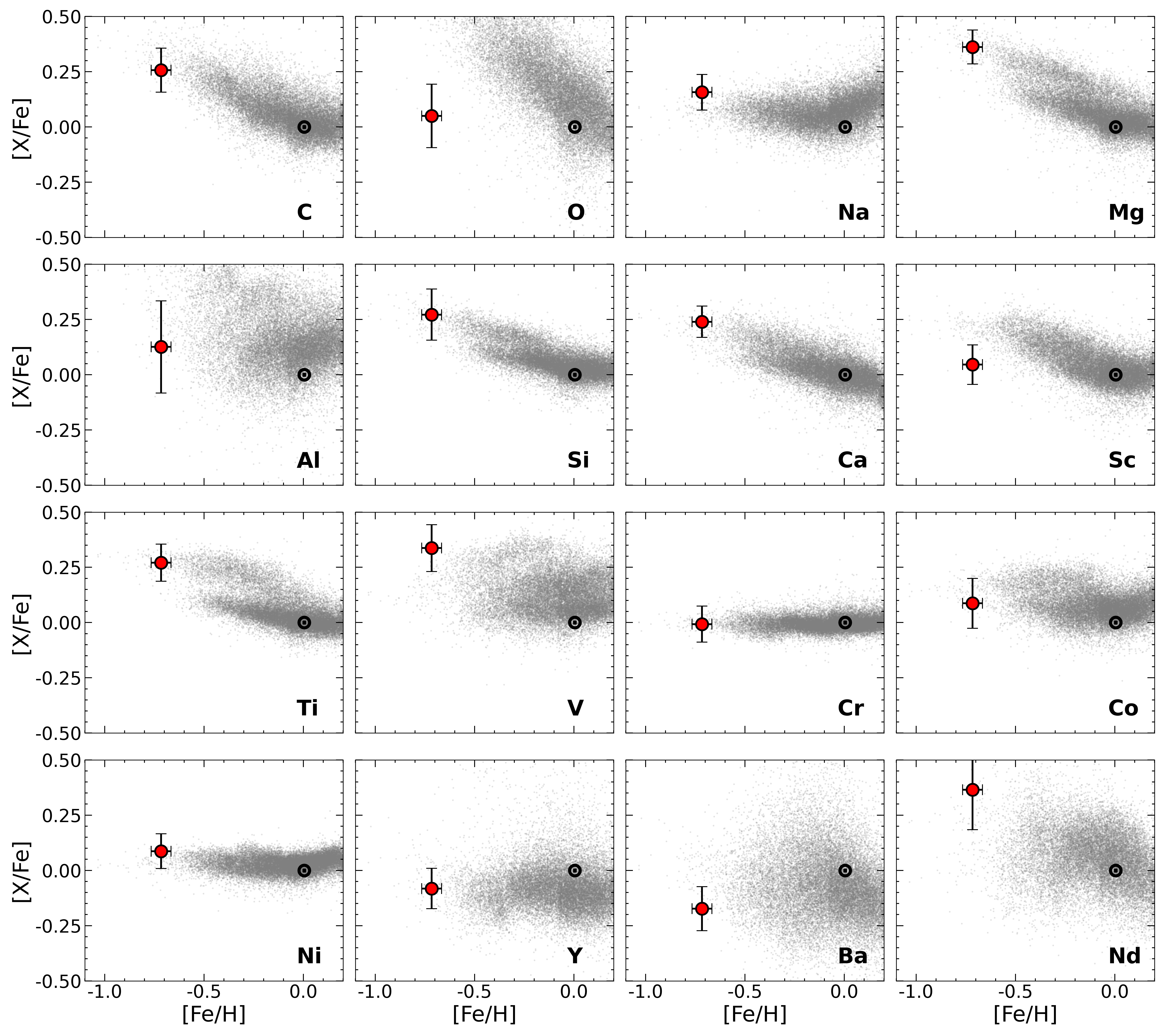}
    \caption{Abundance pattern of TOI-7169.  Each of the 16 panels shows the abundance ratio of one element with respect to iron, plotted against metallicity.
    In each panel, the derived abundance for TOI-7169 is displayed as a red circle, the solar abundance is indicated by the black sun symbol, and abundances of main-sequence stars from GALAH DR4 \citep{buder25} are shown as small gray dots.  The TOI-7169 abundances generally agree with the extrapolated trends from GALAH, with the exception of oxygen, which has a significantly lower abundance than expected for a star of its metallicity.}\label{fig:abundances}
\end{figure*}

The elemental abundances of TOI-7169 are listed in Table~\ref{tab:abunds} and displayed in Fig.~\ref{fig:abundances} along with a comparison sample of stars with broadly similar stellar parameters from the GALAH survey \citep{buder25}.  The GALAH stars are almost all at higher metallicities, but the TOI-7169 abundances are mostly near the values expected if the GALAH abundance trends were extrapolated to $\feh = -0.72$.  The most notable exception is oxygen, where TOI-7169 has an abundance consistent with solar and typical metal-poor stars have $\mathrm{[O/Fe]} > 0.5$.  GALAH and our study measure oxygen from the same features and both apply NLTE corrections, so any systematic offsets should be small.  The low oxygen abundance of TOI-7169 is in contrast to its elevated ratios of the other $\alpha$-elements (e.g., Mg, Si, and Ca).  The star is also mildly enriched in carbon ($\mbox{[C/Fe]} = 0.26$) and moderately enriched in the neutron-capture elements past the second $r$-process peak; lighter neutron-capture species are not enhanced.  The combination of a low oxygen abundance and higher carbon abundance produces a slightly supersolar value of $\mbox{[C/O]} = 0.18$.  The corresponding carbon to oxygen ratio is $n_{\mathrm{C}}/n_{\mathrm{O}} = 0.89$.  This abundance pattern is typical for stars of similar age and metallicity, but TOI-7169 has a higher C/O ratio than most planet-hosting stars \citep[e.g.,][]{hinkel14,pavlenko19,dm21}.  Because carbon and oxygen are two of the primary molecule-forming species in planetary atmospheres, their abundance ratio may have an observable impact on atmospheric chemistry and planet properties \citep[e.g.,][]{madhusudhan11,alidib14,helling17}.

\begin{deluxetable*}{llccccccc}[th!]
\tablecolumns{8}
\tabletypesize{\footnotesize}
\tablecaption{\label{tab:abunds}Stellar Abundances.}
\tablehead{ $Z$ & Element & $\log{\epsilon}$ & $\Delta\log{\epsilon}$ & [X/H] & $\sigma_{\text{[X/H]}}$ & [X/Fe] & $\sigma_{\text{[X/Fe]}}$ & number of lines }
\startdata
 6 &  \ion{C}{1} & 8.00 & 0.08 & $-0.46$ & 0.09 & \phs0.26 & 0.10 & 2 \\
 8 &  \ion{O}{1} & 8.05 & 0.13 & $-0.64$ & 0.13 & \phs0.08 & 0.14 & 3 \\
11 & \ion{Na}{1} & 5.66 & 0.06 & $-0.56$ & 0.06 & \phs0.16 & 0.08 & 2 \\
12 & \ion{Mg}{1} & 7.20 & 0.05 & $-0.35$ & 0.06 & \phs0.37 & 0.08 & 5 \\
13 & \ion{Al}{1} & 5.84 & 0.20 & $-0.59$ & 0.20 & \phs0.13 & 0.21 & 2 \\
14 & \ion{Si}{1} & 7.07 & 0.02 & $-0.44$ & 0.04 & \phs0.28 & 0.06 & 11 \\
20 & \ion{Ca}{1} & 5.82 & 0.04 & $-0.48$ & 0.05 & \phs0.24 & 0.07 & 9 \\
21 & Sc$^{(a)}$ & 2.47 & 0.06 & $-0.67$ & 0.07 & \phs0.05 & 0.09  & 14 \\
22 & Ti$^{(a)}$ & 4.53 & 0.04 & $-0.45$ & 0.06 & \phs0.27 & 0.08  & 22 \\
23 & \ion{V}{1} & 3.52 & 0.05 & $-0.38$ & 0.09 & \phs0.34 & 0.11  & 5 \\
24 & Cr$^{(a)}$ & 4.93 & 0.05 & $-0.69$ & 0.06 & \phs0.03 & 0.08  & 11 \\
25 & \ion{Mn}{1} & 4.36 & 0.04 & $-1.06$ & 0.07 &  $-$0.34 & 0.09 & 5 \\
26 & Fe$^{(a)}$ & 6.74 & 0.03 & $-0.72$ & 0.05 &  ...  &  ...     & 77 \\
27 & \ion{Co}{1} & 4.31 & 0.08 & $-0.63$ & 0.09 & \phs0.09 & 0.10 & 4 \\
28 & \ion{Ni}{1} & 5.57 & 0.05 & $-0.63$ & 0.06 & \phs0.09 & 0.08 & 11 \\
29 & \ion{Cu}{1} & 3.48 & 0.14 & $-0.70$ & 0.15 & \phs0.02 & 0.16 & 2 \\
30 & \ion{Zn}{1} & 4.09 & 0.08 & $-0.47$ & 0.09 & \phs0.25 & 0.10 & 2 \\
38 & \ion{Sr}{1} & 2.13 & 0.08 & $-0.70$ & 0.10 & \phs0.02 & 0.11 & 2 \\
39 &  \ion{Y}{2} & 1.41 & 0.06 & $-0.80$ & 0.08 &  $-$0.08 & 0.09 & 4 \\
40 & \ion{Zr}{2} & 2.02 & 0.10 & $-0.57$ & 0.10 & \phs0.15 & 0.11 & 2 \\
56 & \ion{Ba}{2} & 1.38 & 0.07 & $-0.89$ & 0.09 &  $-$0.17 & 0.10 & 3 \\
58 & \ion{Ce}{2} & 1.20 & 0.10 & $-0.38$ & 0.10 & \phs0.34 & 0.11 & 3 \\
60 & \ion{Nd}{2} & 1.07 & 0.17 & $-0.35$ & 0.17 & \phs0.37 & 0.18 & 2 \\
62 & \ion{Sm}{2} & 0.72 & 0.07 & $-0.23$ & 0.08 & \phs0.49 & 0.09 & 3 \\
\enddata
\tablenotetext{}{$^{(a)}$ Weighted average of the neutral and ionized species.}
\end{deluxetable*}

\subsection{Derivation of Planetary Properties}
\label{sec:exofast}

\begin{figure*}
    \centering
    \includegraphics[width=0.995\textwidth]{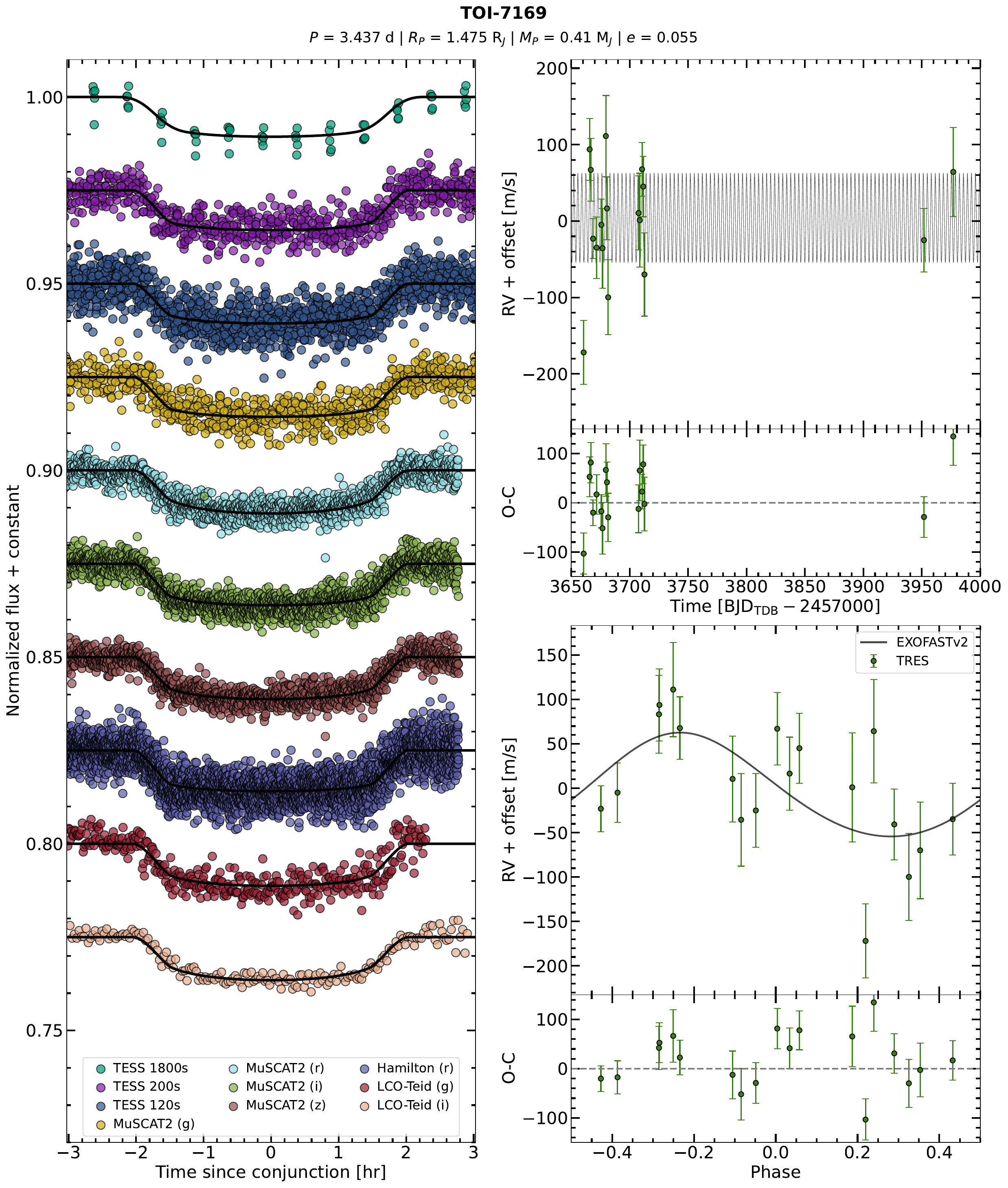}
    \caption{EXOFASTv2 fit results for TOI-7169.  The left panel displays the stacked TESS transits and the full transits observed from the ground along with the best-fit model of each.  The panel in the upper right plots the TRES radial velocities and the best-fit orbit model, with residuals below.  Note that the two velocity measurements from 2016 are not plotted in the upper right panel so that the bulk of the data remains visible.  The panel in the lower right shows the phase-folded radial velocities.}\label{fig:exofast}
\end{figure*}

We carried out a combined fit of the TESS light curve, ground-based photometry sequences, and TRES RV measurements using EXOFASTv2 \citep{eastman13,Eastman:2019}, following the general approach of \citet{rodriguez21} and \citet{schulte24}.  Because EXOFASTv2 does not use $\alpha$-enhanced isochrones that would be appropriate for TOI-7169, we did not include stellar modeling in the fit, but imposed the $1\sigma$ parameter ranges for the stellar mass, radius, temperature, metallicity, and age from the spectroscopic analysis (\S~\ref{sec:hoststar}) as Gaussian priors.  The EXOFASTv2 fit results are listed in Table~\ref{tab:372048733} and compared to the observed transit light curves and radial velocity measurements in Fig.~\ref{fig:exofast}.

\providecommand{\bjdtdb}{\ensuremath{\rm {BJD_{TDB}}}}
\providecommand{\tjdtdb}{\ensuremath{\rm {TJD_{TDB}}}}
\providecommand{\feh}{\ensuremath{\left[{\rm Fe}/{\rm H}\right]}}
\providecommand{\teff}{\ensuremath{T_{\rm eff}}}
\providecommand{\teq}{\ensuremath{T_{\rm eq}}}
\providecommand{\ecosw}{\ensuremath{e\cos{\omega_*}}}
\providecommand{\esinw}{\ensuremath{e\sin{\omega_*}}}
\providecommand{\msun}{\ensuremath{\,M_\Sun}}
\providecommand{\rsun}{\ensuremath{\,R_\Sun}}
\providecommand{\lsun}{\ensuremath{\,L_\Sun}}
\providecommand{\mj}{\ensuremath{\,M_{\rm J}}}
\providecommand{\rj}{\ensuremath{\,R_{\rm J}}}
\providecommand{\me}{\ensuremath{\,M_{\rm E}}}
\providecommand{\re}{\ensuremath{\,R_{\rm E}}}
\providecommand{\fave}{\langle F \rangle}
\providecommand{\fluxcgs}{10$^9$ erg s$^{-1}$ cm$^{-2}$}
\providecommand{\densitycgs}{g cm$^{-3}$}
\startlongtable
\begin{deluxetable*}{lcccccccccccc}
\tablecaption{Median values and 68\% confidence intervals for TOI-7169~b, created using EXOFASTv2 commit number 6ba004d4}
\tablehead{\colhead{~~~Parameter} & \colhead{Description} & \multicolumn{11}{c}{Values}}
\startdata
\smallskip\\\multicolumn{2}{l}{Planetary Parameters:}&\smallskip\\
~~~~$P$\dotfill &Period (days)\dotfill &$3.4373125^{+0.0000020}_{-0.0000019}$\\
~~~~$R_P$\dotfill &Radius (\rj)\dotfill &$1.475\pm0.029$\\
~~~~$M_P$\dotfill &Mass (\mj)\dotfill &$0.41\pm0.14$\\
~~~~$T_C$\dotfill &Observed time of conjunction$^{1}$ (\bjdtdb)\dotfill &$2460560.15083^{+0.00061}_{-0.00040}$\\
~~~~$a$\dotfill &Semi-major axis (AU)\dotfill &$0.04280^{+0.00016}_{-0.00017}$\\
~~~~$i$\dotfill &Inclination ($\degr$)\dotfill &$84.62\pm0.26$\\
~~~~$e$\dotfill &Eccentricity \dotfill &$0.055^{+0.11}_{-0.041}$\\
~~~~$\omega_*$\dotfill &Argument of periastron ($\degr$)\dotfill &$-6^{+71}_{-110}$\\
~~~~$T_{\rm eq}$\dotfill &Equilibrium temperature$^{2}$ (K)\dotfill &$1631^{+19}_{-20}$\\
~~~~$\tau_{\rm circ}$\dotfill &Tidal circularization timescale (Gyr)\dotfill &$0.030^{+0.013}_{-0.012}$\\
~~~~$K$\dotfill &RV semi-amplitude (m s$^{-1}$)\dotfill &$60\pm20$\\
~~~~$R_P/R_*$\dotfill &Radius of planet in stellar radii \dotfill &$0.10116\pm0.00054$\\
~~~~$a/R_*$\dotfill &Semi-major axis in stellar radii \dotfill &$6.14^{+0.12}_{-0.11}$\\
~~~~$\delta$\dotfill &$\left(R_P/R_*\right)^2$ \dotfill &$0.01023\pm0.00011$\\
~~~~$\tau$\dotfill &In/egress transit duration (days)\dotfill &$0.02237^{+0.00093}_{-0.00092}$\\
~~~~$T_{14}$\dotfill &Total transit duration (days)\dotfill &$0.16882^{+0.00085}_{-0.00084}$\\
~~~~$b$\dotfill &Transit impact parameter \dotfill &$0.573^{+0.021}_{-0.023}$\\
~~~~$b_S$\dotfill &Eclipse impact parameter \dotfill &$0.567^{+0.024}_{-0.026}$\\
~~~~$T_{S,14}$\dotfill &Total eclipse duration (days)\dotfill &$0.1681^{+0.0043}_{-0.0054}$\\
~~~~$\delta_{S,2.5\mu m}$\dotfill &BB eclipse depth at 2.5$\mu$m (ppm)\dotfill &$537\pm21$\\
~~~~$\delta_{S,5.0\mu m}$\dotfill &BB eclipse depth at 5.0$\mu$m (ppm)\dotfill &$1382^{+34}_{-35}$\\
~~~~$\delta_{S,7.5\mu m}$\dotfill &BB eclipse depth at 7.5$\mu$m (ppm)\dotfill &$1817^{+37}_{-38}$\\
~~~~$\rho_P$\dotfill &Density (\densitycgs)\dotfill &$0.159^{+0.055}_{-0.054}$\\
~~~~$logg_P$\dotfill &Surface gravity (cgs)\dotfill &$2.67^{+0.13}_{-0.18}$\\
~~~~$\fave$\dotfill &Incident flux (\fluxcgs)\dotfill &$1.591^{+0.080}_{-0.083}$\\
~~~~$T_S$\dotfill &Observed time of eclipse$^{1}$ (\bjdtdb)\dotfill &$2460558.48^{+0.28}_{-0.10}$\\
~~~~$T_P$\dotfill &Time of periastron (\tjdtdb)\dotfill &$2460559.46^{+1.2}_{-0.51}$\\
~~~~$M_P/M_*$\dotfill &Mass ratio \dotfill &$0.00044\pm0.00015$\\
~~~~$P_{T,G}$\dotfill &A priori transit prob \dotfill &$0.1801^{+0.0077}_{-0.0063}$\\
\smallskip\\\multicolumn{2}{l}{Wavelength Parameters:}& \\
~~~~$u_{1}$(g')\dotfill &Linear limb-darkening coefficient \dotfill & $0.484\pm0.028$\\
~~~~$u_{1}$(r')\dotfill &Linear limb-darkening coefficient \dotfill & $0.263\pm0.031$ \\
~~~~$u_{1}$(i')\dotfill &Linear limb-darkening coefficient \dotfill & $0.337\pm0.028$\\
~~~~$u_{1}$(z')\dotfill &Linear limb-darkening coefficient \dotfill & $0.177\pm0.039$\\
~~~~$u_{1}$(TESS)\dotfill &Linear limb-darkening coefficient \dotfill & $0.250\pm0.023$\\
~~~~$u_{2}$(g')\dotfill &Quadratic limb-darkening coefficient \dotfill & $0.232\pm0.031$ \\
~~~~$u_{2}$(r')\dotfill &Quadratic limb-darkening coefficient \dotfill & $0.299^{+0.033}_{-0.032}$ \\
~~~~$u_{2}$(i')\dotfill &Quadratic limb-darkening coefficient \dotfill & $0.294\pm0.031$ \\
~~~~$u_{2}$(z')\dotfill &Quadratic limb-darkening coefficient \dotfill & $0.283^{+0.044}_{-0.043}$ \\
~~~~$u_{2}$(TESS)\dotfill &Quadratic limb-darkening coefficient \dotfill & $0.280^{+0.024}_{-0.023}$ \\
\smallskip\\\multicolumn{2}{l}{Telescope Parameters:}&TRES\smallskip\\
~~~~$\gamma_{\rm rel}$\dotfill &Relative RV offset (m s$^{-1}$)\dotfill &$-57^{+15}_{-14}$\\
~~~~$\sigma_J$\dotfill &RV jitter (m s$^{-1}$)\dotfill &\phs$43^{+18}_{-16}$\\
\enddata
\label{tab:372048733}
\tablenotetext{}{See Table 3 in \citet{Eastman:2019} for a detailed description of all parameters}
\tablenotetext{1}{Time of conjunction is commonly reported as the ``transit time''}
\tablenotetext{2}{Assumes no albedo and perfect redistribution}
\end{deluxetable*}

From the transit depth and the stellar radius, we measure a planetary radius of $1.475\pm0.029~R_{\mathrm{J}}$.  Based on the TRES radial velocities, the radial velocity semi-amplitude of TOI-7169 is $60\pm20$~m~s$^{-1}$, and the orbit is consistent with being circular.  (We also ran the EXOFAST fit with the eccentricity held fixed at zero, finding that all fit parameters agree with the free-eccentricity fit within the uncertainties.)  The planet mass is $0.41 \pm 0.14~M_{\mathrm{J}}$, making TOI-7169~b an inflated Jovian planet with a density of $0.159^{+0.055}_{-0.054}$~g~cm$^{-3}$.  The time baseline of the TRES spectra spans $\sim9$~yr, providing significant sensitivity to long-period stellar or substellar companions.  We carried out an initial EXOFAST fit allowing for a linear trend in the radial velocities, which produced a best fit consistent with no trend (as is visually evident from the data in Fig.~\ref{fig:exofast} and Table~\ref{tab:vels}).  The final fit described above was therefore run with the trend fixed to zero.

\section{Discussion}
\label{sec:discussion}

\subsection{Comparison to Exoplanet Population and Occurrence Rates}

As discussed in \S~\ref{sec:intro}, confirmed planets have been identified down to metallicities of $\feh \approx -0.7$, with some candidate planets at $\feh = -0.8$, $-0.9$, and below.  The best candidates for genuine exoplanets in this metallicity range include Kapteyn's~ Star~b and c at $\feh = -0.89$ \citep[][although see \citealt{bortle21}]{anglada14}, HD~11755~b at $\feh = -0.74$ \citep{lee15}, BD+03~2562 at $\feh = -0.71$ \citep{villaver17}, and 24~Boo~b at $\feh \approx -0.7$ \citep{takarada18}.   However, none of these planets transit their host stars, so their composition and atmospheres are not observable.  In this metallicity range, the transiting sample consists primarily of small planet candidates, with host star metallicities that have not been spectroscopically confirmed.  The only transiting giant planets in the literature at $\feh < -0.5$ are WASP-98~b ($\feh = -0.60$ from \citealt{hellier14} and $\feh=-0.49 \pm 0.10$ from \citealt{mancini16}) and WASP-112~b ($\feh = -0.64$ from \citealt{anderson14} and $\feh = -0.54 \pm 0.05$ from \citealt{ferreira25}).  TOI-7169 is therefore likely the most metal-poor host of a transiting giant planet currently known.

\citet{boley21} determined an upper limit on the occurrence rate of short-period giant planets of 0.18\%\ for stars at $\feh < -0.6$.  This result suggests that a sample of $\gtrsim550$ metal-poor stars would need to be searched in order to find a planet like TOI-7169~b.  From Table~\ref{tab:372048733}, the transit probability for TOI-7169~b is 0.18, which increases the required search sample to $\gtrsim3100$ metal-poor stars.  The Gaia~DR3 catalog contains $\sim360,000$ stars with $G < 12.5$ and spectrophotometric metallicities $\mbox{[M/H]} < -0.6$.  Presuming that TESS has observed most of these stars, the \citet{boley21} limit indicates that up to $\sim100$ similar planets could be detectable. 

In addition to their variation with metallicity, \citet{zink23} found that planet occurrence rates are also a function of Galactic oscillation height (the maximum distance away from the Galactic plane that a star's orbit reaches), in the sense that stars found farther above or below the plane are less likely to host planets.  There is not yet any consensus on the origin of this correlation, but suggestions include an increase in the formation rate of systems with tightly packed inner planets with time \citep{ballard24}, decreased accretion rates from the interstellar medium for protoplanetary disks outside the Galactic midplane \citep{winter24}, and the intense ultraviolet radiation fields experienced by disks that formed $\gtrsim10$~Gyr ago when the local star formation rate was much higher \citep{hl25}.
As shown in \S~\ref{sec:orbit}, despite its low metallicity, TOI-7169 has a maximum $z$-height of 200~pc, whereas \citeauthor{zink23} did not detect a measurable decrease in occurrence rates until $z \gtrsim 500$~pc.  If the \citet{zink23} measurements apply to giant planets (their sample was only statistically significant for small planets), then TOI-7169~b may fit in with the idea that stars closer to the Galactic plane are more likely to host planets, independent of their metallicities.

\subsection{Is a Low Host Star Metallicity Associated with Low Planet Density?}

\begin{figure}
    \centering
    \includegraphics[width=\linewidth]{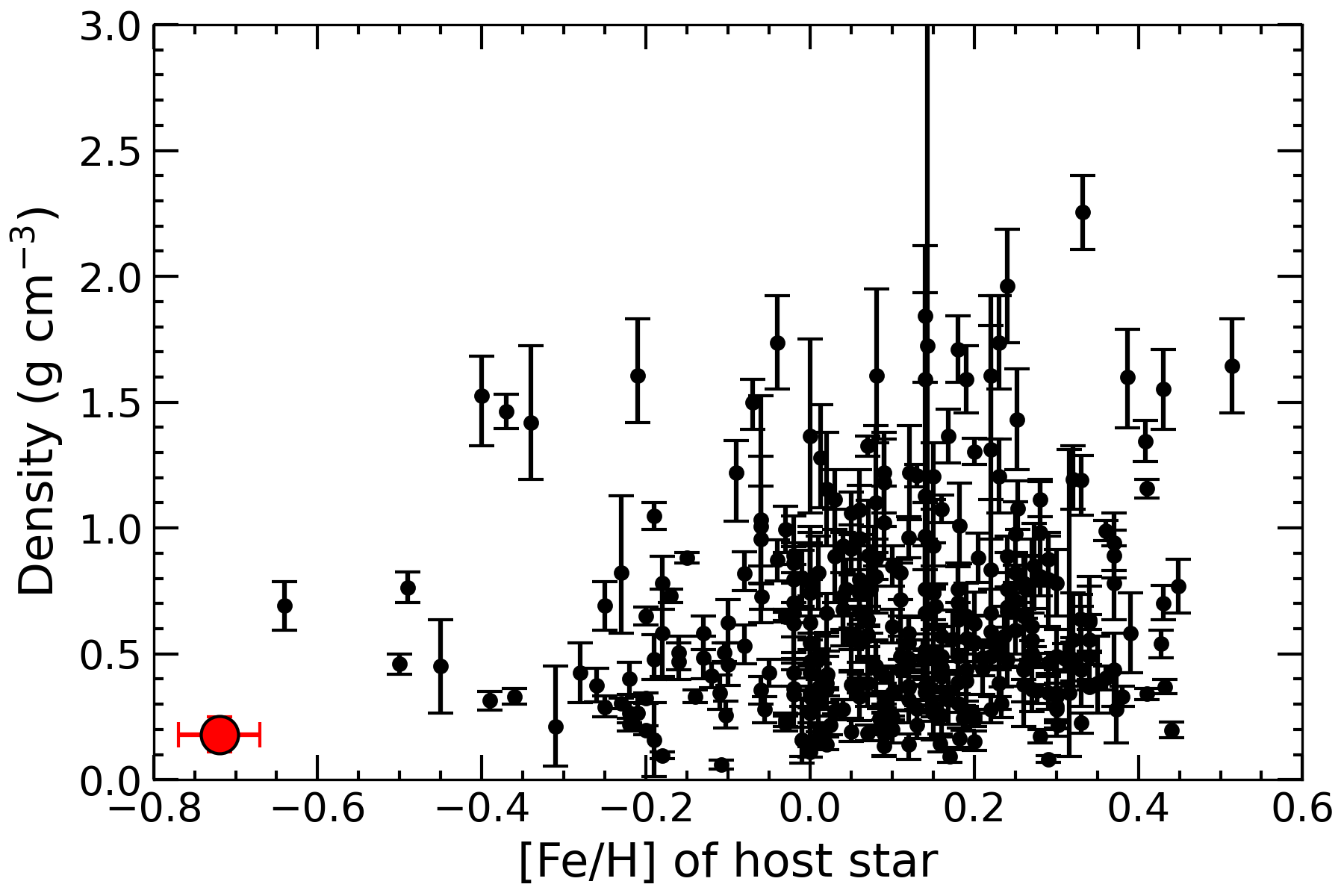}
    \caption{Giant planet densities as a function of host star metallicity.  Data are taken from the TEPCat database \citep{southworth11}.  We plot planets with radii between 1~$R_{\mathrm{J}}$ and 1.5~$R_{\mathrm{J}}$ and masses less than 2.0~$M_{\mathrm{J}}$.  TOI-7169~b is displayed as a large red circle.  Although it may appear by eye that the planets around the most metal-poor host stars are biased to lower densities, the large number of low density planets with metal-rich hosts that are obscured in the bulk of the distribution mean that this tendency is not statistically significant.\label{fig:density}}
\end{figure}

Given the low density of TOI-7169~b, it is tempting to conclude that a low host star metallicity is related to the formation of less dense planets.   Other giant planets around metal-poor stars, such as WASP-98~b \citep{hellier14} and WASP-112~b \citep{anderson14}, also have densities substantially lower than that of Jupiter, supporting this idea.  In addition, \citet{behmard25} showed that for rocky planets, there is a significant anti-correlation between density and [Mg/Fe].  The low density and high Mg abundance of TOI-7169~b are consistent with this trend extending to larger planets as well.  However, WASP-37~b, WASP-46~b, and WASP-163~b represent counterexamples of giant planets with Jupiter-like densities despite host star metallicities of $\feh < -0.3$  \citep{simpson11,ciceri16,barkaoui19}.  The full population of exoplanets with similar sizes from the database of \citet{southworth11} shows no correlation of density with host star metallicity, and a K-S test indicates that the densities of giant planets around metal-poor host stars are consistent with being drawn from the same distribution as the densities of giant planets with metal-rich hosts (see Fig.~\ref{fig:density}).  Thus, TOI-7169~b may be inflated by the hot Jupiter radius anomaly rather than a metallicity-specific mechanism, although models do suggest that a lower bulk metallicity results in larger radii \citep{chachan25}.

\subsection{Orbit in the Milky Way}
\label{sec:orbit}

\begin{figure*}
    \centering
    \includegraphics[width=0.99\textwidth]{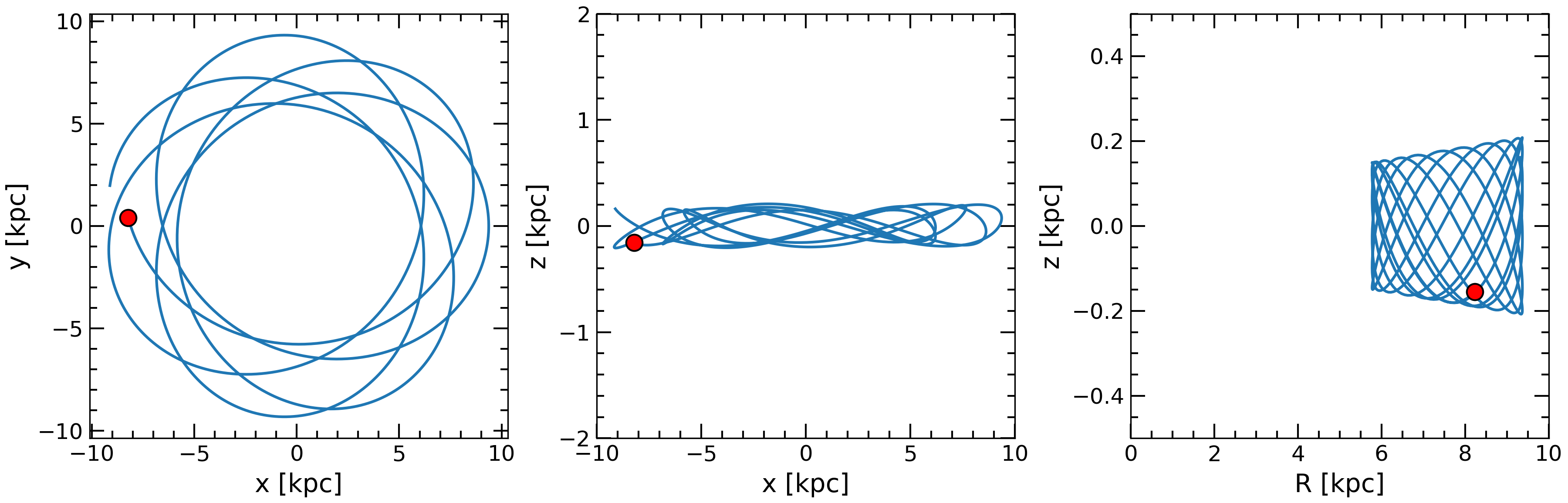}
    \caption{Galactic orbit of TOI-7169 over the past 1~Gyr.  The left panel shows the orbit in the Cartesian $x$--$y$ plane, the middle panel in the Cartesian $x$--$z$ plane, and the right panel in cylindrical coordinates.  Note the change in y-axis scale between the middle and right panels.  In each panel, the red circle indicates the current position of TOI-7169.  As expected for an old star, the orbit is somewhat eccentric, but it remains confined within a narrow range around the Galactic plane.}\label{fig:orbit}
\end{figure*}

Using the distance, proper motion, and velocity of TOI-7169 from Gaia~DR3, we computed the orbit of the star around the Galaxy with the \texttt{gala} package \citep{gala}.  The orbital path over the last 1~Gyr is displayed in Fig.~\ref{fig:orbit}.  The orbit is confined within the Milky Way disk, with a moderate eccentricity of $e = 0.24$.  The pericenter of the orbit is 5.8~kpc and the apocenter is 9.4~kpc, so TOI-7169 remains within $\sim2$~kpc of the solar circle throughout its orbit.  Surprisingly for such an old star, the orbit is confined within 200~pc of the Galactic plane.  TOI-7169 was likely born close to the time that the Milky Way disk first formed \citep[e.g.,][]{bk22}.

\begin{figure}
    \centering
    \includegraphics[width=\linewidth]{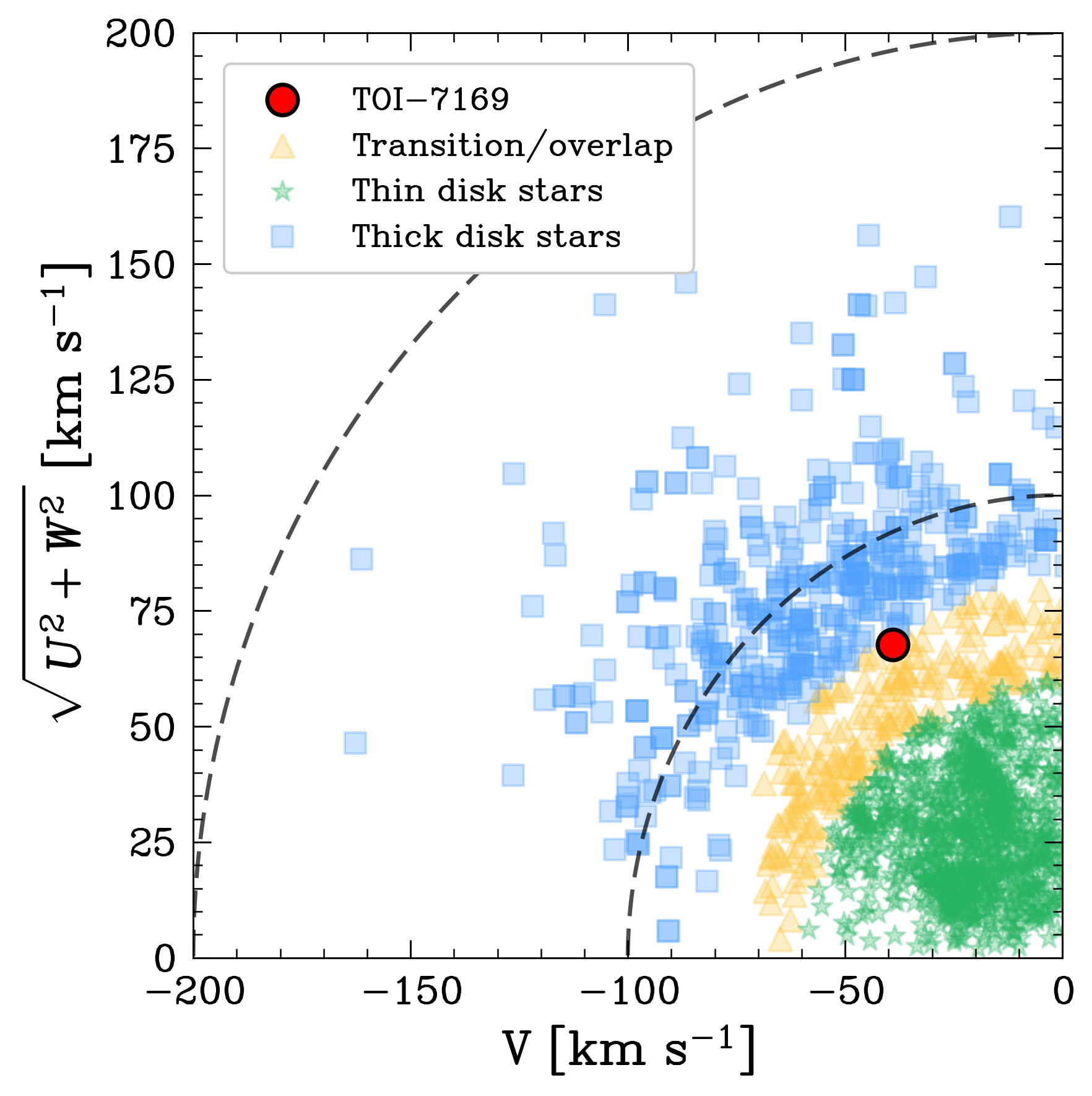}
    \caption{Toomre diagram for TOI-7169. The dashed curves indicate constant total velocities of 100~km~s$^{-1}$ and 200~km~s$^{-1}$. Green stars, yellow triangles, and blue squares represent the thin disk ($v_{\rm{tot}} \leq 60$~km~s$^{-1}$), transition/overlap ($60 < v_{\rm{tot}} \leq 80$~km~s$^{-1}$), and thick disk ($80 < v_{\rm{tot}} \leq 180$~km~s$^{-1}$) populations, respectively, following the definitions of \citet{Feltzing:2003A&A...397L...1F}. Data taken from the NASA Exoplanet Archive \citep{2013PASP..125..989A}.}\label{fig:toomre}
\end{figure}

There are innumerable analyses of the structure of the Milky Way disk in the literature, which reach different quantitative conclusions about the scale heights of the thin and thick disks based on the stellar tracer used, the age and metallicity (and [$\alpha$/Fe]) ranges considered, and the parameterization adopted for the structure of each component.  However, broadly speaking, the thin disk scale height is generally found to be $\lesssim300$~pc \citep[e.g.,][]{bhg16} whereas the thick disk scale height is $\gtrsim500$~pc \citep[e.g.,][]{robin14,bhg16}.  For old and $\alpha$-enhanced stars, the scale height locally is again at least $\sim500$~pc \citep[e.g.,][]{mackereth17,mv18,imig25,xiang25}.  The median absolute deviation of the TOI-7169 orbit from the Galactic midplane of 127~pc is therefore more consistent with the thin disk.  Using Gaia~DR3 kinematics, the $U$, $V$, and $W$ space velocities of TOI-7169 are $-67.70$, $-39.00$, and $-1.49$~km~s$^{-1}$, respectively.  The $U$ and $V$ velocities and the total space velocity ($\sqrt{U^2 + V^2 + W^2} = 78.1$~km~s$^{-1}$) are consistent with the thick disk (see Fig. \ref{fig:toomre}), but as noted above, the behavior in the $z$ direction is much more characteristic of the thin disk.  Using the method of \citet{bensby14}, the ratio of the thick disk membership probability to that of the thin disk is 0.11.

\subsection{Spectroscopy Metrics}

Many of the atoms and molecules found in giant planet atmospheres are composed of metals \citep[e.g.,][]{ws15,dyrek24,hoeijmakers24,gressier25,louie25,ss25}.  It is likely that the abundance of these species and the overall atmospheric chemistry is affected by the availability of metals during planet formation.  TOI-7169~b may therefore represent an intriguing contrast with the hot Jupiters around primarily metal-rich host stars that have been studied recently with JWST and other telescopes.  In the framework of \citet{kempton18}, we compute a transmission spectroscopy metric (TSM) of 166, making the planet a quite feasible target for spectroscopy with JWST.  Objects with similar TSM values that have already been studied with JWST include 
GJ~9827~d \citep{piaulet24}, LTT-9779~b \citep{radica24}, and Kepler-51~d \citep{libbyroberts26}.
The emission spectroscopy metric (ESM) of TOI-7169~b of 54 is somewhat less favorable, but still in the range of other JWST targets whose emission spectra have been detected (e.g., WASP-17~b; \citealt{gressier25} and 55~Cnc~e; \citealt{hu24,patel24}).



\subsection{Implications for Hot Jupiter Formation and Evolution}

Theoretical models suggest that planet formation may be inhibited below a critical metallicity.  The small number of known planets in the metal-poor regime, however, has made it difficult to test this idea quantitatively.  Based on the timescale for dust grain settling in a protoplanetary disk, \citet{jl12} found a minimum metallicity of $\feh_{\mathrm{crit}} \simeq -1.5 + \log{(r/1~\mathrm{AU})}$.  At the current semi-major axis of 0.04~AU for TOI-7169~b, this limit translates to $\feh = -2.9$, which is far below the actual metallicity of TOI-7169.  Plugging the measured metallicity into this framework, the planet would need to have formed at a distance of no more than 6~AU from the host star.  If the system contains additional planets at longer periods (see below), the formation scenario constraints may become stronger.

With an age of 12~Gyr, TOI-7169~b is one of the oldest known planets \citep[e.g.,][]{campante15,grieves22,weeks25}.  Since various formation pathways for hot Jupiters have been proposed, the properties of this system may provide insight into the formation of other hot Jupiters.  One interesting question that may be investigated with future observations is whether TOI-7169 hosts an outer companion that could have been responsible for high-eccentricity migration of TOI-7169~b \citep[e.g.,][]{dj18}.  The lack of an RV signal with an amplitude above $\sim200$~m~s$^{-1}$ over the 9 years covered by our spectroscopy rules out stellar or brown dwarf companions with periods less than 1--2 decades.  Wider companions in the G and K star range should also have been detected in our existing data, but a low-mass M dwarf or brown dwarf in a long period orbit could have escaped detection.  At shorter periods ($\sim$1--10~yr), additional giant planets cannot be excluded without an improved RV data set.  Long-term radial velocity monitoring, higher contrast imaging, and searches for transiting objects with longer periods could detect or further constrain the existence of such a companion.  Astrometric measurements from the fourth Gaia data release could also be sensitive to a companion, although the reduced unit weight error (RUWE) and astrometric excess noise in DR3 do not currently provide any evidence that the motion of TOI-7169 deviates from that of a single star.


\section{Conclusions}
\label{sec:conclusions}

As part of a program aiming to find transiting planets around metal-poor stars, we
have presented the discovery and characterization of the giant planet TOI-7169~b.  TOI-7169 was identified as exhibiting the signal of a potential transiting planet by the TESS team.  As part of a dedicated search for planets orbiting metal-poor host stars, we selected it from the list of TESS Objects of Interest as a candidate based on metallicity estimates from Gaia spectrophotometry.

We obtained ground-based imaging and spectroscopy of TOI-7169 that confirmed both the existence of the planet TOI-7169~b and the metal-poor nature of the host star.  TOI-7169 is a slightly evolved early G star with a mass of $0.885\pm0.010~M_\odot$, a metallicity of $\feh = -0.72 \pm 0.05$, and an age of $12.3 \pm 0.6$~Gyr.  Based on this metallicity measurement, TOI-7169 is the most metal-poor star known to host a transiting giant planet.  The star is $\alpha$-enhanced, at $[\alpha/\mathrm{Fe}]= 0.29$, but despite its old age and low metallicity, its Galactic orbit identifies it as a member of the thin disk population, reaching a maximum distance from the Galactic plane of 200~pc.

We determined that TOI-7169~b is a giant planet with a radius of $1.475\pm0.029~R_{\mathrm{J}}$, a mass of $0.41 \pm 0.14 ~M_{\mathrm{J}}$, and an orbital period of 3.44~d.  The planetary density is quite low ($0.159^{+0.055}_{-0.054}$~g~cm$^{-3}$) and its orbit is consistent with being circular.  The available data do not provide any evidence of a longer period companion in the system that could be responsible for the migration of TOI-7169~b to its current location, but we cannot rule out additional planets at the Jupiter or super-Jupiter mass scale in orbits beyond $\sim1$~AU.

The unique combination of the large radius of the planet, the low metallicity of the host star, and its relatively bright magnitude ($J = 11.20$) make TOI-7169 a promising target for transmission spectroscopy.  Comparison of the atmospheric structure and composition with those of other planets orbiting more metal-rich stars may provide insight into how planetary atmospheres are affected by the availability of heavy elements.  Moreover, the survival of TOI-7169~b for more than 12~Gyr will constrain the evolutionary processes that shape planets over long time scales.

\begin{acknowledgments}
We thank Allyson Bieryla, Madeleine McKenzie, Emily Pass, and Shreyas Vissapragada for advice and assistance during this work.  We thank the anonymous referee for suggestions that clarified the paper.

This research has made use of NASA's Astrophysics Data System Bibliographic Services.

Funding for the TESS mission is provided by NASA's Science Mission Directorate. 

This research has made use of the Exoplanet Follow-up Observation Program (ExoFOP; DOI: 10.26134/ExoFOP5) website, which is operated by the California Institute of Technology, under contract with the National Aeronautics and Space Administration under the Exoplanet Exploration Program.

K.A.C. acknowledges support from the TESS mission via subaward s3449 from MIT.  A.C. is supported by the Brinson Foundation through a Brinson Prize Fellowship grant.  Funding for K.B. was provided by the European Union (ERC AdG SUBSTELLAR, GA 101054354).   We acknowledge financial support from the Agencia Estatal de Investigaci\'on of the Ministerio de Ciencia
e Innovaci\'on MCIN/AEI/10.13039/501100011033 and the ERDF “A way of making Europe” through projects PID2021-125627OB-C32 and PID2024-158486OB-C32.  F. M. acknowledges the financial support from the Agencia Estatal de
Investigaci\'{o}n del Ministerio de Ciencia, Innovaci\'{o}n y
Universidades (MCIU/AEI) through grant PID2023-152906NA-I00.  This work was partly supported by JSPS KAKENHI Grant Numbers JP24H00017, JP24K00689 and JSPS Bilateral Program Number JPJSBP120249910.

This paper has made use of the Python package GaiaXPy, developed and maintained by members of the Gaia Data Processing and Analysis Consortium (DPAC), and in particular, Coordination Unit 5 (CU5), and the Data Processing Centre located at the Institute of Astronomy, Cambridge, UK (DPCI).

This article is based on observations made with the MuSCAT2 instrument, developed by ABC, at Telescopio Carlos Sánchez operated on the island of Tenerife by the IAC in the Spanish Observatorio del Teide.

Some of the observations in this paper made use of the NN-EXPLORE Exoplanet and Stellar Speckle Imager (NESSI). NESSI was funded by the NASA Exoplanet Exploration Program and the NASA Ames Research Center. NESSI was built at the Ames Research Center by Steve B.\ Howell, Nic Scott, Elliott P.\ Horch, and Emmett Quigley.

This work makes use of observations from the LCOGT network. Part of the LCOGT telescope time was granted by NOIRLab through the Mid-Scale Innovations Program (MSIP). MSIP is funded by NSF.

TRAPPIST is funded by the Belgian Fund for Scientific Research (Fond National de la Recherche Scientifique, FNRS) under the grant PDR T.0120.21. TRAPPIST-North is a project funded by the University of Liege (Belgium), in collaboration with Cadi Ayyad University of Marrakech (Morocco). 
M.G. and E.J. are FNRS-F.R.S. Research Directors.


\end{acknowledgments}



\facilities{TESS, Gaia, FLWO:1.5m, TRAPPIST, LCOGT, MuSCAT2}

\software{AstroImageJ \citep{Collins:2017}, astropy \citep{astropy}, EXOFASTv2 \citep{eastman13,Eastman:2019}, gala \citep{gala}, IRAF, keplersplinev2 \citep{Vanderburg:2014}, kmpfit \citep{KapteynPackage}, lightkurve \citep{lightkurve}, matplotlib
\citep{Hunter:2007}, MOOG \citep{Sneden:1973PhDT.......180S}, Qoyllur-quipu \citep{Ramirez:2014A&A...572A..48R}, TAPIR \citep{Jensen:2013}, uberMS \citep{pass25}}

\bibliographystyle{aasjournalv7}

\end{document}